\documentclass[prd,tightenlines,nofootinbib,superscriptaddress]{revtex4}

\usepackage{amsfonts,amssymb,amsthm,bbm}
\usepackage{amsmath}

\usepackage{color,psfrag}
\usepackage[dvips]{graphicx}

\usepackage{subcaption}
\usepackage{graphicx}
\usepackage{xcolor}
\usepackage{tabularray}
\usepackage{physics}
\usepackage{csquotes}
\usepackage{booktabs}
\usepackage{float}

\usepackage{hyperref}  

\newcommand{\be}{\begin{equation}}
\newcommand{\ee}{\end{equation}}
\newcommand{\beq}{\begin{eqnarray}}
\newcommand{\eeq}{\end{eqnarray}}
\newcommand{\bes}{\begin{eqnarray}}
\newcommand{\ees}{\end{eqnarray}}

\newcommand{\lp}{\left(}
\newcommand{\rp}{ \right)}
\newcommand{\lc}{\left[}
\newcommand{\rc}{\right]}

\newcommand{\f}{\frac}

\def\nn{\nonumber}
\def\pp{\partial}

\def\eps{\epsilon}
\def\om{\omega}

\begin{document}

\title{Highly damped Quasi-Normal Modes of a Loop Quantum Black Hole}

\author{\bf Clara Montagnon}\email{clara.montagnon@ens-lyon.fr}
\affiliation{Univ Lyon, ENS de Lyon, CNRS, Laboratoire de Physique, F-69342 Lyon, France}

\begin{abstract}

\medskip

We compute asymptotic Quasi-Normal Mode (QNM) frequencies -- \textit{i.e.} frequencies with a very large Imaginary part -- of a Loop Quantum Gravity inspired Black Hole. The deformations from the Schwarzschild Black Hole are encoded via two parameters: the minimal area gap $a_0$ and the polymeric deformation parameter $P$. In this study, we focus on the effect of the latter one, $P$, on the highly-damped part of QNM spectra. We consider both spin 0 and spin 2 test-field perturbations on the Black Hole as proper gravitational perturbations cannot be performed on an effective model. 

We use an analytical method of computation of QNMs referred to as the monodromy technique, which allows us to compute the asymptotic behaviour of QNMs. We found interesting oscillating behaviour in both the Real part and the Imaginary part of the QNMs, where the oscillation period varies with the polymeric deformation parameter $P$. We compare these analytical predictions to numerical results obtained thanks to the Continued fraction method. Even though the latter does not converge for QNMs with a very large Imaginary part, the numerical results are in rather good agreement with the monodromy prediction.

\end{abstract}

\maketitle

{\small \tableofcontents }

\newpage

\section*{\uppercase{Introduction}}

Until now, the theoretical predictions extracted from General Relativity are correctly matching and explaining every experimental result. We can mention for example the dynamics of the Hulse-Taylor binary pulsar system \cite{Hulse:1974eb,Weisberg:1981bh}, the cosmological observations of the PLANCK collaboration \cite{Planck:2018nkj} and the detection of Gravitational Waves (GWs) a decade ago \cite{LIGOScientific:2016aoc}. 
The waves were emitted during the merger of two Black Holes (BHs) and their measurement now allows to test GR in the regime of strong gravity. The LIGO-Virgo-KAGRA collaboration collected data for years since the first detection and all of it correctly matches with the GR predictions, up to the current measurement precision \cite{LIGOScientific:2016lio,LIGOScientific:2019fpa,LIGOScientific:2020tif}. The latter is expected to increase, as well as the detection rate and the range of frequencies spanned thanks to the expected future GW detectors: LISA \cite{LISA:2017pwj,LISACosmologyWorkingGroup:2019mwx,LISA:2024hlh}, Einstein Telescope \cite{Punturo:2010zz,ET:2019dnz,Chiummo:2023kxt} and Cosmic Explorer \cite{Dwyer_2015,Evans:2021gyd}.
In the future, GR will therefore either be validated at an even greater scale or face experimental discrepancies.

In order to compare the GWs measurements to GR, it is important to model the different phases of GW waveforms emitted during a Binary BH merger. We can distinguish three. The first one is the inspiral, during which the two BHs rotate around each other. Then, as they get closer and closer to each other, the GW amplitude increases until the merger phase. Following this, only one BH remains and it emits GWs in order to settle down into a stationary state.  
This last phase is referred as the ringdown, and its signal consist in a superposition of damped sinusoids, along with a GW burst right after the merger and a power-law tail at the right end. The damped sinusoids are characterised by complex frequencies being called the Quasi-Normal Modes (QNMs) \cite{Kokkotas:1999bd,Nollert:1999ji,Berti:2009kk,Konoplya:2011qq,franchini2023testing}. The measurement of QNMs via the detections of GWs emitted by Binary BHs is now being referred to as \textit{Black Hole Spectroscopy} and became of great interest as it appears to be a great tool for testing of GR. Indeed, as the QNMs solely depend on the BH parameters, their detection gives hope of finding quantum signatures \cite{Berti:2005ys,Berti:2018vdi,Maselli:2019mjd,LIGOScientific:2016lio,Ghosh:2021mrv}.

The breakdown of GR in the center of BHs, among other failures of the theory itself, led to the search for alternative theories of gravity. We can distinguish two main directions: either extending GR or developing a mathematical formalism to describe new microscopic space-time structures. 
An example of the latter is Loop Quantum Gravity (LQG), which is a non-perturbative and background independent approach to quantizing GR. Even though it stands as a promising candidate for a theory of Quantum Gravity, it remains pretty challenging to extract predictions directly from the full theory. 
Effective models of BHs are being build using concepts, tools and methods from LQG, and are expressed as quantum corrected Schwarzschild solutions. 

The Modesto BH \cite{Modesto:2008im}, one of the first effective BH model derived from LQG, is characterised by two parameters. The first one is the area gap $a_0$, corresponding to the smallest quantum of area and naturally resolving the BH singularity in the Modesto metric. This parameter is then expected to be very small: $a_0\ll r_s$, and it can then naturally be neglected in some calculations. Nevertheless, we specifically studied the QNMs dependence in this area gap parameter (setting $P$ to zero) in a previous paper \cite{Livine:2024bvo}. Our analysis showed that the very low damped QNMs exhibit small dependence in $a_0$, while mid-damped QNMs are more affected and especially display a crossing of the imaginary axis. This could be a sign of oscillating pattern, but our numerical code had not been able to go further. 
\\
The second parameter is the polymeric deformation function $P$, which modifies the BH structure by splitting the Schwarzschild radius $r_s$ into two radius $r_+(P)$ and $r_-(P)$, where $r_+$ is now the event horizon and $r_-$ is an inner Cauchy horizon. The analysis of astrophysical data allowed to constrained the possible physical values of the polymeric function: $P<6.17\times10^{-3}$ at 95$\%$ confidence level \cite{Zhu:2020tcf,Yan:2022fkr,Liu:2023vfh}.
Our previous analysis has notably demonstrated that the fundamental QNM is deformed up to the third decimal when $P\sim10^{-3}$, compared to the Schwarzschild one. Then we studied the deformations of the Schwarzschild spectrum more generally, going until about $-20i$, the maximum we could reach with our numerical code. It appeared that the deformed spectra looked very alike the Schwarzschild one, the absolute real part being positively shifted and the imaginary gap evolving as about 0.5+$P$. These observations motivated the work towards a more precise description of the asymptotic part. 


The first step has then be to adapt the monodromy technique to the test-field perturbations of the Modesto BH. It turns out that the procedure is more delicate in the case of $a_0\ne0$, so we will focus only on perturbations with $P\ne0$ and set $a_0=0$.
One could expect the monodromy technique to result in vertical asymptotes, as it is the case for the Schwarzschild BH and given what our previous analysis suggested. Yet, it turns out that the structure of the metric is similar to the one of the charged Reissner-Nordström BH and so is the result: we find QNM spectra oscillating both in the Real part and the Imaginary part, with a great variation of patterns according to the value of $P$ and of the angular momentum parameter $\ell$. As we will see, these special patterns were not visible for the ranges of values of $P$ we studied in our previous paper. 
Our upgraded numerical code now allows us to confirm the existence of this oscillating pattern \textit{via} the Continued fraction method, even though it remains challenging to compute highly-damped modes as the continued fraction does not converge properly. The monodromy technique then appears as a great complement to the Continued fraction method in the study of QNMs on the hole complex plane. The values of $P$ for which the oscillations are visible (by continuity, the oscillations of $P\rightarrow0$ have a period $T\rightarrow \infty$) are about $P>0.1$, which is four orders of magnitude above the astrophysical constraint. Yet, we study in this project the whole range of values for P: $0<P<1$, in a aim of better understanding the structure and properties of this case.

We will start by reminding the geometry of the Modesto BH in the subcase $a_0=0$, before giving the equations for scalar and gravitational test-field perturbations. Our work on the BCL BH \cite{Arbey:2025ses} demonstrated that spin 2 test-field QNMs can be very different from the physical gravitational ones, especially in the highly-damped regime. Our findings regarding gravitational test-field perturbations are therefore not intended to make physical predictions, but rather constitute a theoretical mathematical study.
We will then apply the Continued fraction method to these cases, along with the Monodromy technique. The results arising from both computations accompanied by comparisons will be given in section \ref{sec_modesto2_results}.

\section{Effective Loop Quantum Black with simplified geometry}

As mention before, we will restrict to the case $a_0=0$ for two reasons. First, $a_0$ is the scaled version of LQG's area gap $\Delta$: $a_0=\Delta/8\pi$ -- which corresponds to the smallest quantum of area predicted by LQG and is thus expected to be of Planck order --. It is then natural to neglect it. This is especially justified by our previous analysis, which showed that $a_0$ had to be larger than $10^{-2}$ to induce a significant deviation on the QNMs spectrum. Moreover, the structure of the elements involved in the monodromy computation make the procedure more involved as we explain in appendix \ref{appendix_mono_Modesto_a0_case}, and we have not been able to pursuit it.

The metric description given by Modesto in \cite{Modesto:2008im} for his stationary and spherically symmetric effective LQG BH corresponds to
\be 
\dd s^2 = -f(r)\dd t^2 + \frac{\dd r^2}{g(r)} + h(r)\dd \Omega^2,
\label{Modesto_new_metric}
\ee
where, in the case $a_0=0$, the metric functions read
\be 
f(r)=\frac{(r-r_+)(r-r_-)(r+r_0)^2}{r^4}, \quad g(r) = \frac{(r-r_+)(r-r_-)}{(r+r_0)^2}, \quad h(r)=r^2.
\label{metric_P}
\ee
We recall that $r_+$ corresponds to the outer event horizon radius (reducing to the Schwarzschild radius $r_s$ in the Schwarzschild limit), while $r_-$ is an inner Cauchy radius (vanishing in the Schwarzschild limit). They are related to the polymeric deformation function $P$ via
\be 
r_+=\frac{r_s}{(1+P)^2}, \quad r_-=\frac{r_s P^2}{(1+P)^2}, \quad r_0=\frac{r_s P}{(1+P)^2} = \sqrt{r_+r_-},
\ee
while $P$ is defined in terms of the Barbera-Immirizi parameter $\gamma$, a constant from the LGQ theory, via
\be 
P = \frac{\sqrt{1+\varepsilon^2}-1}{\sqrt{1+\varepsilon^2}+1}, \quad \varepsilon = \gamma \delta,
\ee
and where $\delta$ is called the polymeric parameter. It is an adimensional effective parameter which is a priori resulting from the coarse-graining of the fundamental Planck scale dynamics. It is then a free parameter and it has been constrained as we mentioned earlier \cite{Zhu:2020tcf,Yan:2022fkr,Liu:2023vfh}: $\delta<0.67$ at 95$\%$ level, hence the constraint on the polymeric function $P<6.17\times10^{-3}$. We will consider values larger than this bound for our study, willing to explore the new structures and symmetries arising in the hole range of possible values for $P$. We will then consider $0<P<1$, as $1$ is an natural upper-bound imposed by its definition.

\section{Scalar and gravitational test-field perturbations}

The Modesto BH being an effective BH model derived from LQG concepts, proper metric perturbations cannot be performed as there is not any equivalent of the Einstein equations available for the description of the modified dynamics. As an alternative, we choose to study spin 0 and 2 test-field perturbations with the aim of analysing the deformations arising from the modifications of the BH model. As mentioned before, the spin 2 test-field perturbations do not take into account the modified dynamics from LQG, yet we study them as a mathematical toy with the aim of getting a better understanding of the structure of QNM spectra exhibiting deformations. 

\begin{figure}[!h]
    \centering
    \begin{subfigure}[b]{0.49\textwidth}
    \centering  \includegraphics[width=0.95\textwidth]{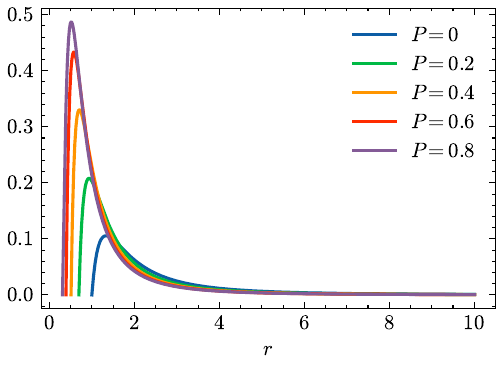}
    \caption{$s=0, \, \ell=0$}
    \label{}
    \end{subfigure}
    \hfill
    \begin{subfigure}[b]{0.49\textwidth}
    \centering
    \includegraphics[width=0.95\textwidth]{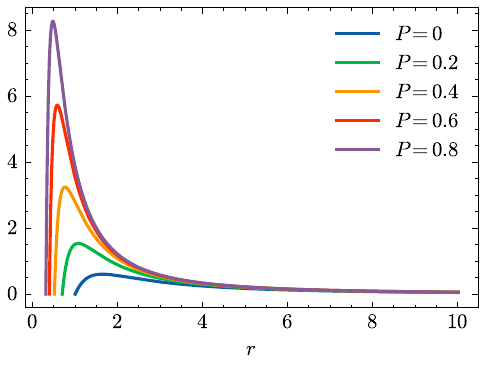}
    \caption{$s=2, \, \ell=2$}
    \label{}
    \end{subfigure}
    \caption{Evolution of the effective potential describing spin 0 and 2 test-field perturbations on the Modesto BH with $a_0=0$. The polymeric deformation parameter $P$ spans between 0 and 0.8. The case $P=0$ corresponds to the Schwarzschild BH.}
    \label{plot_potentials_Modesto_P}
\end{figure}

Given that the Modesto BH is a static and spherically symmetric metric, we can directly use the master equation for the field function $\Psi(r)$
\be
\frac{\dd^2 \Psi}{\dd x^2}+[\omega^2-V_s(r)]\Psi = 0,
\label{master_eq_Modesto_P}
\ee
where the effective potentials $V_s(r)$ are given by the general formulas for scalar and test-field spin 2 perturbations -- which can be found in the paper \cite{Arbey:2021jif} -- applied to the metric functions \eqref{metric_P}:
\begin{subequations}
    \begin{align}
        V_0(r) &= \frac{(r-r_-)(r-r_+)}{r^2}\lc \frac{r_-+r_+}{r^3}+\frac{(r+r_0)^2\lambda-2r_-r_+}{r^4}\rc,\\
        V_2(r) &= \frac{(r-r_-)(r-r_+)}{r^2}\lc \frac{\lambda}{r^2} + \frac{2r_0(\lambda-2)-3(r_-+r_+)}{r^3}+\frac{4r_-r_++r_0^2(\lambda-2)}{r^4}\rc,
        \label{potentials_Modesto_P}
    \end{align}
\end{subequations}
The angular-momentum contribution is given by the parameter $\lambda=\ell(\ell+1)$ and the Schwarzschild limit is recovered by taking $P= 0$. The tortoise coordinate $x$ used to cast the master equation describing the perturbations dynamic into a Schrödinger-like equation is defined via 
\be
\frac{\dd x}{\dd r} = \frac{1}{\sqrt{f(r)g(r)}} = \frac{r^2}{(r-r_+)(r-r_-)}.
\ee
Its integrated expression reads
\be 
x(r) = r+ \frac{r_+^2\log(r-r_+)-r_-^2\log(r-r_-)}{r_+-r_-} + \text{cste}.
\ee 
Finally, we can learn a lot at the behaviour of the spin 0 and 2 test-field potentials \eqref{potentials_Modesto_P} by plotting them for different values of the polymeric function $P$ (we choose to span $P$ from 0 to 0.8), with respect to the radial coordinate $r$. The plots are displayed in figure \ref{plot_potentials_Modesto_P}. 
One can see that the potentials are barrier-type ones, vanishing both at the event horizon $r_+$ and at infinity. The larger the polymeric function $P$ is, the more reduced is the event horizon $r_+$ along with the radial location of the maxima of the effective potential, while the height of the latter is largely increased.

\section{Continued fraction method}

Let us now apply the Continued fraction method developed by Leaver \cite{Leaver:1985ax} by first building an ansatz for the test-field function $\Psi(r)$ which obeys the physically motivated boundary conditions for QNMs.
As we saw in the previous section, the effective potentials vanish both at the horizon $r_+$ and at infinity. Due to the specific shape of the master equation, the asymptotic behaviour of the field function $\Psi$ will then be plane-waves in the tortoise coordinate $x$. We can therefore take a look at the asymptotic first order behaviour of the tortoise coordinate $x$ at the horizon $r_+$ and at infinity:
\begin{subequations}
    \begin{align}
        x = \frac{r_+^2 \log(r-r_+)}{r_+-r_-} + \mathcal{O}(1), \quad (r\rightarrow r_+),\\
        x = r + (r_-+r_+)\log(r) + \mathcal{O}(1), \quad (r\rightarrow \infty).
    \end{align}
\end{subequations}
Knowing that QNMs are required to be ingoing only at the BH event horizon and outgoing only at infinity, we can write the asymptotic behaviour of the field function $\Psi$ in terms of the radial coordinate $r$: 
\begin{subequations}
    \begin{align}
        \Psi \sim e^{-i\omega x} \sim (r-r_+)^{-i\omega \frac{r_+^2}{r_+-r_-}}, \quad (r\rightarrow r_+),\\
        \Psi \sim e^{i\omega x} \sim e^{i\omega r} + r^{i\omega (r_-+r_+)}, \quad (r\rightarrow \infty).
    \end{align}
\end{subequations}
Following Leaver's method \cite{Leaver:1985ax}, we can now formulate an ansatz for the field function $\Psi$ describing the scalar and gravitational test-field perturbations, in term of a power series in the variable $\frac{r-r_+}{r-r_-}$:
\be 
\Psi(r)= e^{i\omega(r-r_+)}(r-r_-)^{i\omega(r_-+r_+)}\lp \frac{r-r_+}{r-r_-}\rp^{-i\omega \frac{r_+^2}{r_+-r_-}} \sum_{n=0}^{\infty}a_n\lp \frac{r-r_+}{r-r_-}\rp^n.
\ee 
In order for this ansatz to be physically well defined, the power series need to respect some conditions. First, we have to require $a_0\neq 0$ and $a_n=0$ for $n<0$ to ensure the QNM boundary condition at horizon. Second, the series $\sum a_n$ must be convergent in order for the ansatz to have a correct behaviour at infinity.

Plugging this ansatz into the second-order differential master equation with the effective potentials $V_0(r)$ and $V_2(r)$ yields recursion equations that must be satisfied by the coefficients $a_n$: 
\begin{equation}
\label{recur_eq}
\left|
    \begin{array}{rcl}
        \alpha_0a_1+\beta_0a_0&=&0
        \,,\\
        \alpha_1a_2+\beta_1a_1+\gamma_1a_0&=&0
        \,,\\
        \alpha_2a_3+\beta_2a_2+\gamma_2a_1+\delta_2a_0&=&0
        \,,\\
        \alpha_n a_{n+1}+\beta_na_n+\gamma_na_{n-1}+\delta_na_{n-2}+\epsilon_na_{n-3}
        &=&0\,, \qquad\textrm{for}\quad n\geq 3.
    \end{array}
\right.
\end{equation}
The five coefficients $\alpha_n, \beta_n, \gamma_n, \delta_n, \epsilon_n$ are all quadratic polynomials in $n$. They also depend on the polymeric deformation function $P$, the frequency $\omega$ and the angular momentum parameter $\lambda$.
\\
They read explicitly for spin $s=0$: 
\begin{subequations}
\begin{align}
\alpha_n &= n^2 (1-P) (1 + P)^5 - 2 i n (1 + P)^2 \omega,
\\
\beta_{n}&= 2 n^2 ( P-1) (1 + P)^5 (1 + P^2) + 2 n \left[ (1-P) (1 + P)^5 (1 + \lambda+2\lambda P+(3+\lambda) P^2) \right.\\& \left. + i (1 + P)^2 (4 - P^2 + P^6) \omega \right] + (P-1) (1 + P)^5 (1 + 3 P^2)\nn\\& - i (1 + P)^2 (4 + 3 P^2 + P^6) \omega - 2 (-4 + P^2 + P^4) \omega^2,\nn
\\
\gamma_{n} &= n^2 (1 - P) (1 + P)^5 (1 + 4 P^2 + P^4) + 2 n (1 + P)^2 \left[ (P-1) (1 + P)^3 (1 + 8 P^2 + 3 P^4) \right. \\& \left. -  i (2 + 6 P^2 - 5 P^4 + 2 P^6 + P^8) \omega \right] - (P-1)(1 + P)^5 (1 + 14 P^2 + 9 P^4 +2\lambda P(1+P)^2) \nn \\& +  2 i (1 + P)^2 (2 - 2 P^2 + P^4) (1 + 8 P^2 + 3 P^4) \omega + (-4 - 12 P^2 + 5 P^4 - 3 P^6 + P^8 + P^{10}) \omega^2,\nn
\\
\delta_n &= 2 n^2 (P-1) P^2 (1 + P)^5 (1 + P^2) + 2 n P^2 (1 + P)^2 \left[ (1 - P) (1 + P)^3 (5 + 7 P^2) \right.\\& \left. + i (3 + 2 P^2 - 3 P^4 + 2 P^6) \omega \right] + P^2 \left[(P-1) (1 + P)^5 (11 + \lambda + 2\lambda P + (25+\lambda) P^2) \right. \nn \\& \left. - i (1 + P)^2 (17 + 7 P^2 (2 - 3 P^2 + 2 P^4)) \omega + 2 (2 + 3 P^2 + 2 P^6 - P^8) \omega^2\right],\nn
\\
\eps_n &= n^2 (1 - P) P^4 (1 + P)^5 + n P^4 \lc 8 (P-1) (1 + P)^5 - 2 i (1 + P)^2 \omega \rc \\&+ P^4 \lc16 (1 - P) (1 + P)^5 + 8 i (1 + P)^2 \omega - (1 + P^2) (1 + P^4) \omega^2 \rc,\nn
\end{align}
\end{subequations}
\ \\ 
and for spin $s=2$:
\begin{subequations}
\begin{align}
\alpha_n &= n^2 (1-P) (1 + P)^5 - 2 i n (1 + P)^2 \omega,
\\
\beta_{n}&= 2 n^2 ( P-1) (1 + P)^5 (1 + P^2) + 2 n \left[ (1-P) (1 + P)^5 (1 + 3 P^2) \right.\\& \left. + i (1 + P)^2 (4 - P^2 + P^6) \omega \right] + (P-1)(1 + P)^5 (\lambda-3 + P(-4+3P+(2+P)\lambda)\nn\\& -i P^2 (1 + P)^2 (3 + P^4)\omega - 2 (-4 + P^2 + P^4) \omega^2,\nn
\\
\gamma_{n} &= n^2 (1 - P) (1 + P)^5 (1 + 4 P^2 + P^4) + 2 n (1 + P)^2 \left[ (P-1) (1 + P)^3 (1 + 8 P^2 + 3 P^4) \right. \\& \left. -  i (2 + 6 P^2 - 5 P^4 + 2 P^6 + P^8) \omega \right] -(1 + P)^5 \lc 3 + P + P^2 (-18 + P (18 + P (-9 + 5 P)))\rc \nn \\& + 2 (P-1) P (1 + P)^7 \lambda +  2 i (1 + P)^2 (2 - 2 P^2 + P^4) (1 + 8 P^2 + 3 P^4) \omega \nn \\&+ (-4 - 12 P^2 + 5 P^4 - 3 P^6 + P^8 + P^{10}) \omega^2,\nn
\\
\delta_n &= 2 n^2 (P-1) P^2 (1 + P)^5 (1 + P^2) - 2 n P^2 (1 + P)^2 \left[ ((-1 + P) (1 + P)^3 (5 + 7 P^2)) \right. \\& \left. + i (3 + 2 P^2 - 3 P^4 + 2 P^6) \omega \right] + P^2 \left[ (P-1) (1 + P)^5 (11 +\lambda + P (-4 + 21 P +(2+P)\lambda)) \right. \nn \\& \left. - i (1 + P)^2 (17 + 7 P^2 (2 - 3 P^2 + 2 P^4)) \omega + 2 (2 + 3 P^2 + 2 P^6 - P^8) \omega^2 \right],\nn
\\
\eps_n &= n^2 (1 - P) P^4 (1 + P)^5 + n P^4 \lc 8 (P-1) (1 + P)^5 - 2 i (1 + P)^2 \omega \rc \\&+ P^4 \lc16 (1 - P) (1 + P)^5 + 8 i (1 + P)^2 \omega - (1 + P^2) (1 + P^4) \omega^2 \rc.\nn
\end{align}
\end{subequations}
\ \\

In order to apply Leaver's method, we first need to reduce this five-terms recurrence relation into a three-terms one which will be cast-able in a continued fraction equation. To this end, we apply a two steps Gaussian reduction. 
It first consists in defining four new recursion coefficients such that
\be
\bar{\alpha}_n= \alpha_n
\,,\quad
\bar{\beta}_n= \beta_n-\frac{\bar{\alpha}_{n-1}}{\bar{\delta}_{n-1}}\epsilon_n,\quad
\bar{\gamma}_n= \gamma_n-\frac{\bar{\beta}_{n-1}}{\bar{\delta}_{n-1}}\epsilon_n,\quad
\bar{\delta}_n= \delta_n-\frac{\bar{\gamma}_{n-1}}{\bar{\delta}_{n-1}}\epsilon_n.
\ee
The five-terms recursion relation is then equivalent to a four-terms recursion relation in these new coefficients:
\be
\bar\alpha_n a_{n+1}+\bar\beta_na_n+\bar\gamma_na_{n-1}+\bar\delta_na_{n-2}=0.
\ee
The same procedure can be applied a second time to define three new recursion coefficients
\be
 \bar{\bar{\alpha}}_n= \bar{\alpha}_n,\quad
 \bar{\bar{\beta}}_n= \bar{\beta}_n-\frac{\bar{\bar{\alpha}}_{n-1}}{\bar{\bar{\gamma}}_{n-1}}\bar{\delta}_n,\quad
 \bar{\bar{\gamma}}_n= \bar{\gamma}_n-\frac{\bar{\bar{\beta}}_{n-1}}{\bar{\bar{\gamma}}_{n-1}}\bar{\delta}_n,
\ee
such that we end up with a three-term recursion relation:
\be
\bar{\bar{\alpha}}_{n}a_{n+1}+\bar{\bar{\beta}}_{n}a_n+\bar{\bar{\gamma}}_{n}a_{n-1}=0.
\ee
The Gaussian procedure is pretty involved. It is therefore not possible to extract analytical expressions of the three final coefficients and the latter then have to be computed numerically for each choice of the parameters. 
The three-terms recurrence relation finally allows to follow Leaver's procedure by transforming it into a continued fraction equation: 
\be
\beta_0-\f{\bar{\bar{\alpha}}_0\bar{\bar{\gamma}}_1}{\bar{\bar{\beta}}_1-\f{\bar{\bar{\alpha}}_1\bar{\bar{\gamma}}_2}{\bar{\bar{\beta}}_2-\f{\bar{\bar{\alpha}}_2\bar{\bar{\gamma}}_3}{\bar{\bar{\beta}}_3-...}}}=0.
\ee 
The QNMs can then be found by computing the roots of this equation and its inversions.

\section{Analytical computation of the asymptotic QNMs using the monodromy technique}

It is known that, for the Schwarzschild BH, the Real part of the QNM frequencies converges towards the constant value $(\ln 3)/4\pi$ while the gap in Imaginary part between two succeeding QNMs converges towards $\frac{1}{2}$: 
\be 
\boxed{4\pi r_s \omega =\ln(3)-i(2n+1)\pi, \ \ \ \ n\in\mathbb{N}.}
\label{Sch_mono}
\ee 
This asymptotic behaviour was probed numerically by Nollert \cite{Nollert:1993zz} before being derived analytically by Molt \cite{Motl:2002hd}.
The gap between the QNMs imaginary parts is supposed to probe the deep quantum regime of BHs. One can see (\cite{Motl:2002hd,Motl:2003cd,Natario:2004jd,Dreyer:2002vy,Konoplya:2011qq}) for reviews. 
Here, we want to use the same analytical technique of computation which was used to determine \eqref{Sch_mono}, referred as the monodromy technique, to probe the asymptotic of the test-field QNM spectrum of the Modesto BH and see how the parameter $P$ impacts it.
\
\\ \\

Let us consider the master equation for spin 0 and 2 test-field perturbations of the Modesto BH, considering $a_0=0$:
\be 
\frac{\dd^2 \Psi}{\dd x^2} + [\omega^2-V_s(r)]\Psi = 0,
\label{mono_Modesto_P_schro}
\ee
along with the effective potentials
\begin{subequations}
    \begin{align}
        V_0(r) &= \frac{(r-r_-)(r-r_+)}{r^2}\lc \frac{r_-+r_+}{r^3}+\frac{(r+r_0)^2\lambda-2r_-r_+}{r^4}\rc,\\
        V_2(r) &= \frac{(r-r_-)(r-r_+)}{r^2}\lc \frac{\lambda}{r^2} + \frac{2r_0(\lambda-2)-3(r_-+r_+)}{r^3}+\frac{4r_-r_++r_0^2(\lambda-2)}{r^4}\rc,
        \label{mono_potentials_Modesto_P}
    \end{align}
\end{subequations}
where the quantities $r_+, \ r_-, \ r_0$ are given in terms of the polymeric deformation function $P$:
\be 
r_+=\frac{r_s}{(1+P)^2}, \quad r_-=\frac{r_s P^2}{(1+P)^2}, \quad r_0=\frac{r_s P}{(1+P)^2} = \sqrt{r_+r_-}.
\ee

The key point of the monodromy technique is to extend the usual physical domain of definition $r_+<r<\infty$ to the whole complex plane. Then, the equation \eqref{mono_Modesto_P_schro} is an ordinary differential equation with regular singular points at $r=0$, $r=r_+$, $r=r_-$ and an irregular singular point at $r=\infty$. Thus, any solution from \eqref{mono_Modesto_P_schro} in the physical region extends to the complex plane, but may be multivalued around the singular points. 
\\ \\

Let us first take a look at infinity. As the effective potential \eqref{mono_potentials_Modesto_P} vanishes at infinity, the wave-function $\Psi$ has a plane-wave behaviour:
\be 
\Psi(x)\overset{\infty}{\sim} A_+e^{i\omega x}+A_-e^{-i\omega x}
\ee 
The boundary condition for QNMs at infinity, stating that the wave must be outgoing at infinity, implies $A_-=0$ and then $\Phi(x)\overset{\infty}{\sim} A_+e^{i\omega x}$.
\\ \\

Now let us consider the behaviour of the wave-function $\Psi$ around the singularity $r=0$. At first order, the effective potentials \eqref{mono_potentials_Modesto_P} reduces to 
\be 
V_s(r) \overset{r=0}{\sim} \lc \lambda +2(s-1)\rc \frac{P^4}{(1+P)^8}\frac{1}{r^6} + \mathcal{O}(1).
\ee 
It is important to notice that this potential vanishes when taking $P$ to 0 and that we do not recover the first order Schwarzschild potential $V^{(\text{Sch})} \overset{r=0}{\sim}\frac{(s^2-1)r_s^2}{r^4}$, which is the one considered in the monodromy technique for the Schwarzschild BH. As we shall see later, this will result in the lack of a well defined limit $P\longrightarrow 0$ for the monodromy result and its prediction is then expected to be less accurate for small $P$. One would need to go beyond the first order to overcome this issue.
\\
The first order of the tortoise coordinate around the singularity $r=0$ write
\be 
x\overset{r=0}{\sim} -\frac{(1+P)^4}{3P^2}r^3.
\label{mono_Modesto_x_at_0}
\ee
Notice once again that the limit $P\longrightarrow0$ is not well defined.\\
The first order effective potential can then be written as a function of $x$:
\be 
V_s(x)=\frac{\lc \lambda+2(s-1)\rc}{9x^2}.
\label{mono_Modesto_pot_x}
\ee 
The equation \eqref{mono_Modesto_P_schro} for spin 0 and 2 test-field perturbations around the Modesto BH can thus be simplified around $r=0$ as
\be 
\frac{\dd^2 \Psi}{\dd x}^2 + \lp\omega^2-\frac{\lc \lambda+2(s-1)\rc}{9x^2}\rp \Psi  =0.
\label{mono_Modesto_P_schro_0}
\ee
It is convenient to rescale the $x$-coordinate into $z=\om x$ and introduce the function $\Phi$ such that $\Psi(x)=\Phi(\omega x)\,{\sqrt{2\pi\omega x}}$. This function now satisfies a straightforward Bessel equation:
\be 
z^2\pp_z^2\Phi+z\pp_z\Phi+\lp z^2-\nu^2\rp\Phi=0,
\qquad\textrm{with}\quad
\nu=\frac{\sqrt{4\lambda+8s+1}}{6}.
\ee 
Thus the general solution of equation \eqref{mono_Modesto_P_schro_0} for the wave function $\Psi(x)$ around the singularity $r=0$ can be written in terms of Bessel functions of the first kind $J_{\pm\nu}$:
\be 
\Psi(x) \overset{r=0}{\sim} B_+\sqrt{2\pi \omega x} \ J_\nu(\omega x) + B_-\sqrt{2\pi\omega x} \ J_{-\nu}(\omega x)
\label{mono_Modesto_P_solB}
\ee 

\begin{figure}[!ht]
    \centering
    \includegraphics[width=0.5\linewidth]{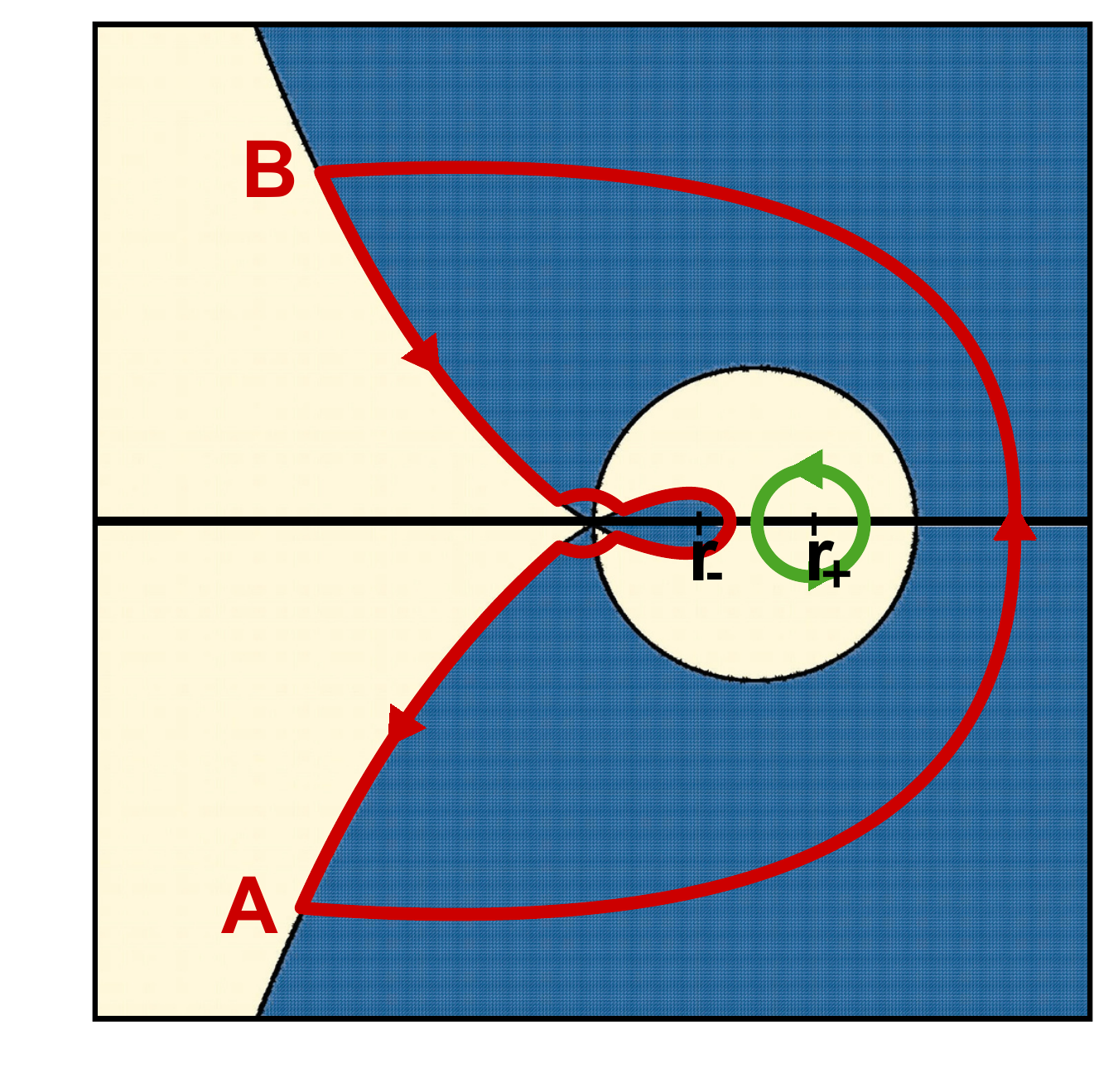}
    \caption{Contour for the calculation of QNM frequencies in the complex $r$ plane. The different colour regions are separated by the associated Stokes lines and the dark blue region corresponds to Re$(x)>0$.}
    \label{Modesto_r_plane}
\end{figure}

The next thing we want to do is matching this solution with the solution at infinity $\Psi(x)\overset{\infty}{\sim} A_+e^{i\omega x}$. In this section we are interested in the study of the asymptotic part of the QNM spectrum. Whereas considering the Schwarzschild BH or the Modesto BH from LQG, one see that going through the spectrum the absolute value of the imaginary part of the QNMs $|\mathrm{Im}(\omega)|$ soon becomes way larger than the real part $\mathrm{Re}(\omega)$. In the asymptotic part of the spectrum, we can then consider $\omega$ to be very large and purely imaginary. We will stick to this hypothesis throughout all the computation.\\
We can now use the following expansion of the Bessel functions
\be
J_\nu(z)\overset{z\gg1}{\sim} \sqrt{\frac{2}{\pi z}} \cos\lp z-\frac{\nu \pi}{2}-\frac{\pi}{4}\rp,
\ee 
and apply it to the solution \eqref{mono_Modesto_P_solB} around $r=0$ for $|\omega x|\gg 1$:
\begin{subequations}
    \begin{align}
        &\Psi(x)\overset{|\omega x|\gg1}{\sim} 2B_+\cos\lp \omega x-\frac{\nu \pi}{2}-\frac{\pi}{4}\rp +  2B_-\cos\lp \omega x+\frac{\nu \pi}{2}-\frac{\pi}{4}\rp,\\
        \iff &\Psi(x)\overset{|\omega x |\gg1}{\sim} \lp B_+ e^{-i\alpha_+}+B_- e^{-i\alpha_-}\rp e^{i\omega x}+ \lp B_+ e^{i\alpha_+} + B_- e^{i\alpha_-}\rp e^{-i\omega x},
\label{BCL_solu1}
    \end{align}
\end{subequations}
with $\alpha_\pm=\frac{\pi}{4}(1\pm 2\nu)$.\\
In order to  match  the solutions at infinity and  around $r=0$, none of the exponentials $e^{\pm i\omega x}$ should dominate each other. 
The matching with the solution at infinity must then be done following the Stokes lines, defined such that $x$ is purely imaginary and then such that $\omega x \in \mathbb{R}$. The Stokes lines are represented in black in the complex r-plane of figure \ref{Modesto_r_plane}. For the calculation of the QNMs frequencies we will follow the red contour, starting at point $B$ where $\mathrm{Re}(\omega x)>0$. Indeed, we have $\mathrm{Im}(\omega)<0$ as we use the convention $e^{-i\omega t}$ for the time evolution, along with $\mathrm{Im}(x)$ being positive at point $B$.
As $\omega x \gg 1$ at point $B$, we can use the solution \eqref{BCL_solu1} and apply the QNM boundary condition at infinity, \textit{i.e.} the condition that the wave must be outgoing at infinity. This give us a first condition: 
\be 
B_+ e^{i\alpha_+} + B_- e^{i\alpha_-} = 0,
\label{mono_Modesto_P_first_cond}
\ee 
and we are left with the following solution for the wave function $\Psi(x)$
\be 
\Psi(x)\overset{\omega x \gg 1}{\sim} \lp B_+ e^{-i\alpha_+}+B_- e^{-i\alpha_-}\rp e^{i\omega x}.
\ee 
In order to rotate from the branch containing the point B to the next one (where the lobe is located), we need to rotate by an angle $-\frac{2\pi}{3}$ in the complex $r$-plane. From equation \eqref{mono_Modesto_x_at_0}, one can see that this is equivalent to rotate by angle $-2\pi$ in the $x$-plane.
Let us apply this to our solution, using the asymptotic expansion for the Bessel function under a $-2\pi$ rotation:
\begin{subequations}
    \begin{align}
        \sqrt{2\pi e^{-2\pi i}\omega x} \ J_{\pm \nu}(e^{-2\pi i}\omega x)&= e^{-\pi i\lp 1\pm 2\nu \rp}\sqrt{2\pi \omega x} \ J_{\pm\nu}(\omega x)\\
        &\sim 2 e^{-4i\alpha_\pm}\cos(\omega x -\alpha_\pm).
    \end{align}
\end{subequations}
The solution on this branch can then be written as:
\begin{subequations}
    \begin{align}
        \Psi(x)&\sim 2 B_+ e^{-4i\alpha_+}\cos(\omega x -\alpha_+) + 2 B_- e^{-4i\alpha_-}\cos(\omega x -\alpha_-)\\
        &= (B_+e^{-3i\alpha_+}+B_- e^{-3i\alpha_-})e^{-i\omega x} + (B_+e^{-5i\alpha_+}+B_- e^{-5i\alpha_-})e^{i\omega x}.
        \label{mono_Modesto_P_solB}
    \end{align}
\end{subequations}
One can see that the behaviour is purely oscillatory. Then following the contour around the lobe circling the inner horizon $r_-$, one travels an additional distance which corresponds to the distance $\delta$ in the $x$-plane such that:
\be 
\delta = -2\pi i \frac{r_-^2}{r_+-r_-}.
\ee 
The solution for $\Psi$ after travelling around the inner horizon $r_-$ can then be written using Bessel functions as in \eqref{mono_Modesto_P_solB}, and taking into account the additional travel distance $\delta$ leads to:
\be 
\Psi(x) \overset{r=0}{\sim} C_+\sqrt{2\pi \omega (x-\delta)} \ J_\nu(\omega (x-\delta)) + B_-\sqrt{2\pi\omega (x-\delta)} \ J_{-\nu}(\omega (x-\delta))
\ee 
As $\omega(x-\delta)$ is negative on this branch, one has to use the asymptotic expansion 
\be
J_\nu (z) \overset{z\ll-1}{\sim} \sqrt{\frac{2}{\pi z}}\cos\lp z+ \frac{\nu\pi}{2} + \frac{\pi}{4} \rp,
\ee 
so that the solution can be expressed as 
\begin{subequations}
    \begin{align}
        \Psi(x) &\sim 2C_+ \cos\lc \om(x-\delta)+\alpha_+ \rc + 2C_- \cos\lc \om(x-\delta)+\alpha_- \rc\\
        &= \lp C_+ e^{-i\alpha_+}e^{i\omega\delta} + C_- e^{-i\alpha_-}e^{i\omega\delta} \rp e^{-i\om x} + \lp C_+ e^{i\alpha_+}e^{-i\omega\delta} + C_- e^{i\alpha_-}e^{-i\omega\delta} \rp e^{i\om x}.
    \end{align}
\end{subequations}
This solution must be matching with the solution \eqref{mono_Modesto_P_solB} we found for $\Psi$ before going around the lobe circling the inner horizon $r_-$. We then have the following conditions:
\begin{subequations}
    \begin{align}
        B_+e^{-3i\alpha_+}+B_- e^{-3i\alpha_-} &= C_+ e^{-i\alpha_+}e^{i\omega\delta} + C_- e^{-i\alpha_-}e^{i\omega\delta}, \label{mono_Modesto_P_second_cond}\\
        B_+e^{-5i\alpha_+}+B_- e^{-5i\alpha_-} &= C_+ e^{i\alpha_+}e^{-i\omega\delta} + C_- e^{i\alpha_-}e^{-i\omega\delta}. \label{mono_Modesto_P_third_cond}
    \end{align}
\end{subequations}
The next step is to do a last rotation to the branch containing the point A by rotating $(x-\delta)$ through an angle of $-2\pi$:
\begin{subequations}
    \begin{align}
        \sqrt{2\pi e^{-2\pi i}\omega (x-\delta)} \ J_{\pm \nu}(e^{-2\pi i}\omega (x-\delta))&= e^{-\pi i\lp 1\pm 2\nu \rp}\sqrt{2\pi \omega (x-\delta)} \ J_{\pm\nu}(\omega (x-\delta))\\
        &\sim 2 e^{-4i\alpha_\pm}\cos(\omega (x-\delta) +\alpha_\pm).
    \end{align}
\end{subequations}
such that the solution on the branch containing the point A reads
\begin{subequations}
    \begin{align}
        \Psi(x) &\sim 2C_+ e^{-4i\alpha_+}\cos\lc \om(x-\delta)+\alpha_+ \rc + 2C_- e^{-4i\alpha_-} \cos\lc \om(x-\delta)+\alpha_- \rc\\
        &= \lp C_+ e^{-5i\alpha_+} + C_- e^{-5i\alpha_-}\rp e^{i\omega\delta} e^{-i\om x} + \lp C_+ e^{-3i\alpha_+} + C_- e^{-3i\alpha_-}\rp e^{-i\omega\delta} e^{i\om x}.
    \end{align}
\end{subequations}
\\

Finally, let us close the contour around $r\sim\infty$, following the red line, where $x\sim r$ and Re$(x)>0$. As Im$(\omega)\ll 0$, the factor $e^{-i\omega x}$ is exponentially  small on this part of the contour, so that only the coefficient of $e^{i\omega x}$ should be taken into account for the calculation. After having completed the red contour, the coefficient $e^{i\omega x}$ is then multiplied by 
\be 
\frac{\lp C_+e^{-3i\alpha_+}+C_- e^{-3i\alpha_-}\rp e^{-i\om \delta}}{B_+e^{-i\alpha_+}+B_- e^{-i\alpha_-}}.
\ee 
Moreover, as $x\overset{r_+}{\sim}\frac{1}{f'(r_+)}\log(r-r_+),$ the counter-clockwise monodromy of $e^{i\omega x}$ is
\be 
e^{i\omega \lp\frac{1}{f'(r_+)}2\pi i\rp}= e^{i\omega \times 2\pi i \frac{r_+^2}{r_-+r_+}}= e^{-2\pi \omega \frac{r_+^2}{r_-+r_+}},
\ee 
so that the counter-clockwise monodromy of the wave function $\Phi$ around this contour is 
\be 
\frac{\lp C_+e^{-3i\alpha_+}+C_- e^{-3i\alpha_-}\rp e^{-i\om \delta}}{B_+e^{-i\alpha_+}+B_- e^{-i\alpha_-}}e^{-2\pi \omega \frac{r_+^2}{r_-+r_+}}.
\ee 
One can distort the contour without any impact on the monodromy value, as long as the distortion does not cross any singularity. Let us then take a contour simply circling the BH event horizon being $r_+$. This contour is drawn in figure \ref{Modesto_r_plane} as the small green circle.
As the effective potentials \eqref{mono_potentials_Modesto_P} vanish around the horizon $r_+$, the wave-function $\Psi$ has a plane-wave shape:
\be 
\Psi(x)\overset{r_+}{\sim} D_+e^{i\omega x}+D_-e^{-i\omega x}.
\ee 
We can apply the QNM boundary condition at the horizon, stating that the wave must be ingoing. This implies $D_+=0$ and then
\be 
\Psi(x)\overset{r_+}{\sim} D_-e^{-i\omega x}.
\ee 
The counter-clockwise monodromy of the wave function $\Psi$ around the green contour is then
\be 
e^{-i\omega \lp\frac{1}{f'(r_+)}2\pi i\rp}= e^{2\pi \omega \frac{r_+^2}{r_-+r_+}}.
\ee 
As the monodromy for the red contour should be equal to the monodromy for the green contour, we obtain a final condition:
\be 
\frac{\lp C_+e^{-3i\alpha_+}+C_- e^{-3i\alpha_-}\rp e^{-2\pi\om \sigma_-}}{B_+e^{-i\alpha_+}+B_- e^{-i\alpha_-}}e^{-2\pi \omega \sigma_+} = e^{2\pi \omega \sigma_+},
\label{mono_Modesto_P_fourth_cond}
\ee 
where 
\be 
\sigma_- = \frac{r_-^2}{r_-+r_+}, \quad \sigma_+ = \frac{r_+^2}{r_-+r_+};
\ee
At the end of the day, we have four conditions, equations \eqref{mono_Modesto_P_first_cond}, \eqref{mono_Modesto_P_second_cond}, \eqref{mono_Modesto_P_third_cond} and \eqref{mono_Modesto_P_fourth_cond}, representing the two QNM boundary conditions at infinity and at the horizon. They form a system of 4 equations with 4 variables $B_\pm, C_\pm$. In order to have non-trivial solutions, the determinant of the system must vanish: 
\be
\left|
\begin{array}{cccc}
e^{i\alpha_+} & e^{i\alpha_-} & 0 & 0\\
e^{-3i\alpha_+} & e^{-3i\alpha_-} &  -e^{-i\alpha_+}e^{2\pi\om \sigma_-} & -e^{-i\alpha_-}e^{2\pi\om \sigma_-}\\
e^{-5i\alpha_+} & e^{-5i\alpha_-} & - e^{i\alpha_+}e^{-2\pi\om \sigma_-} & -e^{i\alpha_-}e^{-2\pi\om \sigma_-}\\
e^{-i\alpha_+}e^{4\pi \omega \sigma_+} & e^{-i\alpha_-}e^{4\pi \omega \sigma_+} & -e^{-3i\alpha_+}e^{-2\pi\om \sigma_-} & -e^{-3i\alpha_-}e^{-2\pi\om \sigma_-}
\end{array}
\right|=0,
\ee 
\be 
\iff \left|
\begin{array}{cccc}
     e^{2i\alpha_+} & e^{2i\alpha_-} & 0 & 0\\
e^{-2i\alpha_+} & e^{-2i\alpha_-} &  e^{4\pi\om \sigma_-} & e^{4\pi\om \sigma_-}\\
e^{-4i\alpha_+} & e^{-4i\alpha_-} & e^{2i\alpha_+} & e^{2i\alpha_-}\\
e^{4\pi \omega \sigma_+} & e^{4\pi \omega \sigma_+} & e^{-2i\alpha_+} & e^{-2i\alpha_-}
\end{array}
\right|=0,
\ee 
\be 
\iff e^{2i\alpha_+} \left|
\begin{array}{ccc}
     e^{-2i\alpha_-} &  e^{4\pi\om \sigma_-} & e^{4\pi\om \sigma_-}\\
e^{-4i\alpha_-} & e^{2i\alpha_+} & e^{2i\alpha_-}\\
e^{4\pi \omega \sigma_+} & e^{-2i\alpha_+} & e^{-2i\alpha_-}
\end{array}
\right|
- e^{2i\alpha_-} \left|
\begin{array}{ccc}
e^{-2i\alpha_+} &  e^{4\pi\om \sigma_-} & e^{4\pi\om \sigma_-}\\
e^{-4i\alpha_+} & e^{2i\alpha_+} & e^{2i\alpha_-}\\
e^{4\pi \omega \sigma_+} & e^{-2i\alpha_+} & e^{-2i\alpha_-}
\end{array}
\right|
=0,
\ee
\be 
 ... \nn 
\ee 
\be 
\iff \boxed{e^{4\pi\om\sigma_+} + \lc 1+2\cos(2\pi \nu)\rc + \lc 2+2\cos(2 \pi \nu)\rc e^{2\pi\om \sigma_-}=0}.
\label{mono_Modesto_P_result}
\ee 
These solutions $\om$ of this equation then give asymptotic QNMs, with proper behaviour at infinity and at the horizon. This equation does not give an explicit expression for the frequencies $\omega$ in terms of the mode number, as it was the case for the Schwarzshild BH. Here, one has to numerically solve this equation on the complex plane in order to obtain the asymptotic QNMs. The results will be given in the following section.

As one can notice, this equation does not reduce to the one for the Schwarzschild BH \eqref{Sch_mono} and instead yields in the limit $r_-\rightarrow 0$:
\be 
e^{4\pi \om r_+} +3 + 4\cos(2\pi\nu) =0.
\ee 
This can be explained by the fact that the first order term of the effective potentials \eqref{mono_potentials_Modesto_P} around $r=0$ vanish. The same issue occurs with the tortoise coordinate $x$ \eqref{mono_Modesto_x_at_0}, changing the topology of the Stokes lines. The limit $r_-\rightarrow 0$ ($P\rightarrow 0$) is then not well defined and we expect the monodromy formula \eqref{mono_Modesto_P_result} to be less accurate for small values of $P$.
\\
\\

Let us recall the general definition of $\nu$ in terms of the spin $s$ of the perturbation field and of the angular momentum parameter $\ell$:
\be 
\nu=\frac{\sqrt{4\lambda+8s+1}}{6}.
\ee
\ \\

In the case of scalar perturbations (spin $s=0$), we then have 
\be 
\nu = \frac{\sqrt{1+4\lambda}}{6},
\ee
and the equation \eqref{mono_Modesto_P_result} reduces to:
\be
\left\{
    \begin{array}{l}
        e^{\frac{4 \pi  r_+^2 \omega }{r_+-r_-}}+3 e^{\frac{4 \pi r_-^2 \omega }{r_+-r_-}}+2 = 0, \quad \quad \text{for \ } \ell=0,2,3,5,6,8,...\\
       e^{\frac{4 \pi r_+^2 \omega}{r_+-r_-} }-1= 0 \quad \quad \text{for \ } \ell= 1,4,7,10,13,....
    \end{array}
    \right.
    \label{mono_eq_s0}
\ee
The first order monodromy computation for scalar perturbations then predicts asymptotic QNMs with vanishing real part for a set of values of the angular momentum parameter $\ell$:
\be 
\om_n = -i\frac{r_+-r_-}{2r_+^2}n, \quad \ell=1+3p, \ p \in \mathbb{N}.
\label{special_Im}
\ee 
The case $\ell=1$ is actually not well defined in this procedure as the first order of the effective potential \eqref{mono_Modesto_pot_x} vanishes for $\ell=1$. We consider this specific case in the appendix \ref{appendix_mono_Modesto_P_case_l1}. We will look for the roots of the first equation in the following section. 
\\ 

In the case of test-field spin 2 perturbations, we have:
\be 
\nu=\frac{\sqrt{17+4\lambda}}{6}.
\ee
The expression of the equation defining the asymptotic modes is then slightly more involved. It reduces to:
\be 
e^{\frac{4 \pi r_+^2 \omega }{r_+-r_-}}+2 e^{\frac{4 \pi r_-^2 \omega }{r_+-r_-}}+2 \cos \left[\pi \frac{\sqrt{4 \lambda+17}}{3}\right] \left(e^{\frac{4 \pi r_-^2 \omega }{r_+-r_-}}+1\right)+1 = 0.
\label{mono_eq_s2}
\ee 

We will compare those results with the ones obtained from the Continued fraction method in the next section.

\section{Analytical and numerical results}
\label{sec_modesto2_results}

\subsection{Monodromy results}

Let us start with the case of scalar perturbations, for which there exist two monodromy equations according to the value of the angular momentum parameter \eqref{mono_eq_s0}. Indeed, all monodromy QNMs with $\ell=1,4,7,10,...$  will have vanishing real parts and an imaginary value determined by \eqref{special_Im}. On the contrary, the monodromy QNMs with $\ell=0,2,3,5,6,...$ will have non-trivial values. 
The numerical procedure for finding the roots of this equation is straightforward, and we easily computed the QNMs frequencies up to $|$Im$(\om)|=500$. 

\begin{figure}
    \hfill
    \centering
    \begin{subfigure}[b]{0.3\textwidth}
    \centering  
    \includegraphics[width=1\textwidth]{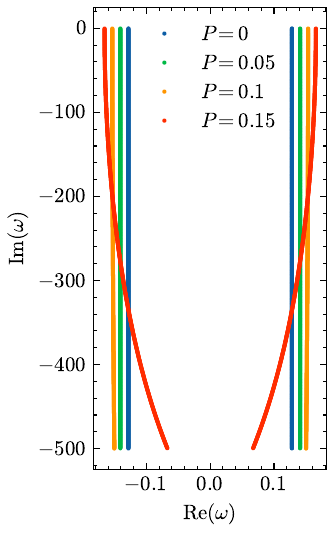}
    \caption{}
    \end{subfigure}
    \hspace{0.2cm}
    \begin{subfigure}[b]{0.5\textwidth}
    \centering  
    \includegraphics[width=1\textwidth]{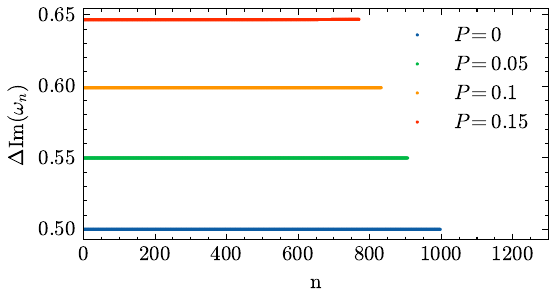}
    \vspace{1.5cm}
    \caption{}
    \end{subfigure}
    \hspace{1cm}
    \caption{(a) Plot in the complex plane of the QNMs values predicted by the monodromy calculation, for scalar perturbations ($s=0$). $P=0$ corresponds to the well-known Schwarzschild case and the other spectra correspond to small deformations. (b) Plot of the gap in imaginary part $\Delta$Im between two successive QNMs.}
    \label{Modesto_mono_s0_small_P}
\end{figure}

The spectra for small deformations in $P$ are displayed in figure \ref{Modesto_mono_s0_small_P}. The spectrum for $P=0$ corresponds to the first order monodromy result of the Schwarzschild BH, and hence consists in a vertical line of real part equal to $\log(3)/4\pi$ and an imaginary gap, defined as 
\be 
\Delta \text{Im} \equiv \text{Im}(\om_n-\om_{n+1}),
\ee
of value 1/2.
One can see that, as $P$ is increased, the real parts and the imaginary gap both increase, but the spectrum keeps a vertical shape, which is consistent with the results of our previous study \cite{Livine:2024bvo}.  In particular, one can see that the monodromy predictions for the imaginary gap \ref{Modesto_mono_s0_small_P} are consistent with our previous observations \cite{Livine:2024bvo} which were based on a computation using the Continued fraction method. However, starting from about $P=0.15$, one can see that the spectrum line is curved. This is the premise of oscillations patterns in the QNM spectra, which appear more and more clearly and intensely as $P$ is large.
One can note that, by continuity, the QNM spectra for $0<P\ll1$ must display a crossing pattern close to infinity.

\begin{figure}[!h]
    \centering
    \begin{subfigure}[b]{0.23\textwidth}
    \centering  
    \includegraphics[width=1\textwidth]{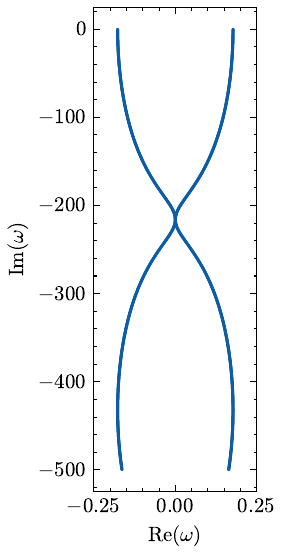}
    \caption{$P=0.2$}
    \end{subfigure}
    \hfill
    \centering
    \begin{subfigure}[b]{0.23\textwidth}
    \centering  
    \includegraphics[width=1\textwidth]{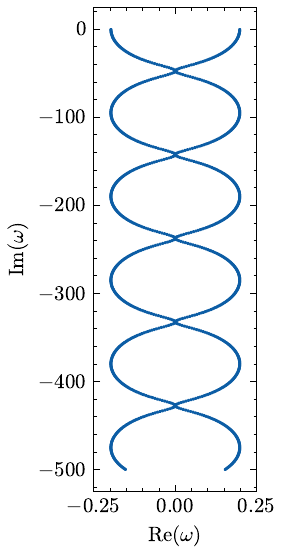}
    \caption{$P=0.3$}
    \end{subfigure}
    \hfill
    \centering
    \begin{subfigure}[b]{0.23\textwidth}
    \centering  
    \includegraphics[width=1\textwidth]{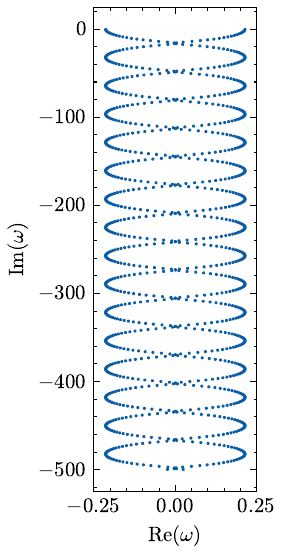}
    \caption{$P=0.4$}
    \end{subfigure}
    \hfill
    \centering
    \begin{subfigure}[b]{0.23\textwidth}
    \centering  
    \includegraphics[width=1\textwidth]{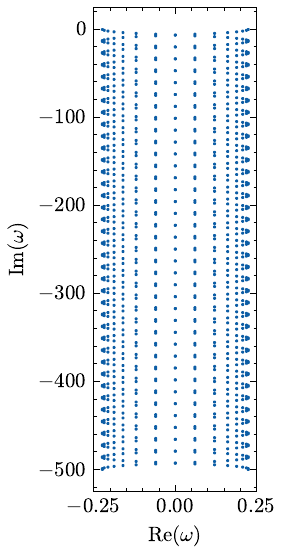}
    \caption{$P=0.5$}
    \end{subfigure}
    \hfill
    \centering
    \begin{subfigure}[b]{0.23\textwidth}
    \centering  
    \includegraphics[width=1\textwidth]{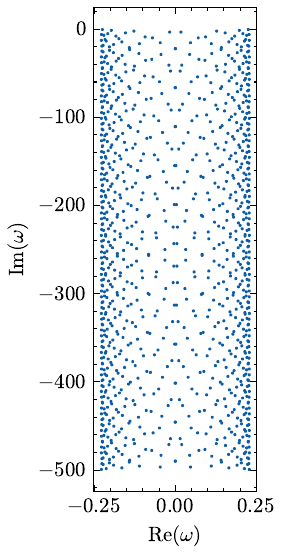}
    \caption{$P=0.6$}
    \end{subfigure}
    \hfill
    \centering
    \begin{subfigure}[b]{0.23\textwidth}
    \centering  
    \includegraphics[width=1\textwidth]{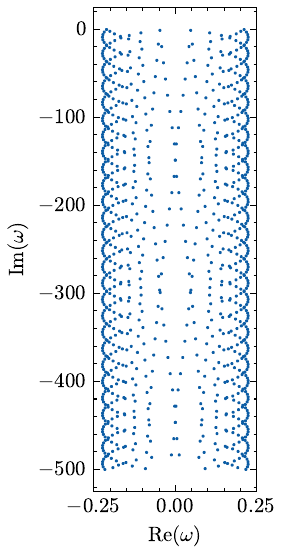}
    \caption{$P=0.7$}
    \end{subfigure}
    \hfill
    \centering
    \begin{subfigure}[b]{0.23\textwidth}
    \centering  
    \includegraphics[width=1\textwidth]{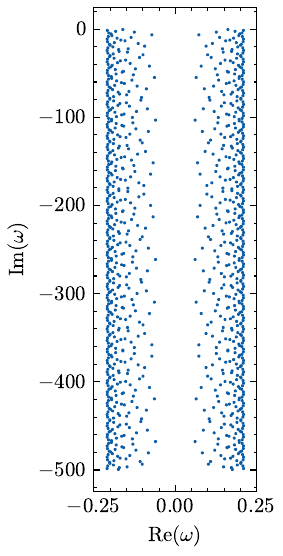}
    \caption{$P=0.8$}
    \end{subfigure}
    \hfill
    \centering
    \begin{subfigure}[b]{0.23\textwidth}
    \centering  
    \includegraphics[width=1\textwidth]{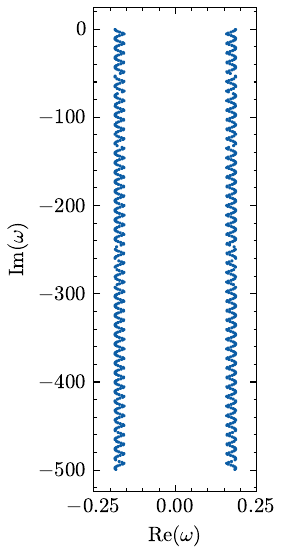}
    \caption{$P=0.9$}
    \end{subfigure}
    \hfill    
    \caption{Plots in the complex plane of the QNMs values predicted by the monodromy calculation, for scalar perturbations ($s=0$) and $\ell=0,2,3,5,6,8,...$.}
    \label{Modesto_mono_plot_s0}
\end{figure}

The QNM spectra for $0.2<P<0.9$ are shown in figure \ref{Modesto_mono_plot_s0}.

One can see that the QNM spectra present strong oscillations patterns in the Real part, with an oscillation period increasing as $P$ enlarges. From $P\sim0.6$, the QNM spectra become highly chaotic and only settle down to clear oscillation patterns around $P=0.9$, where they seem to tighten around one value. One can notice from equation \eqref{mono_eq_s0} that the extremal case $P\sim1$ will be characterised by QNMs with real part equal to Log(3)/4$\pi(r_++r_-)$ and an imaginary gap reaching 1. There are then two extremal and singular cases -- $P=0$ (Schwarzschild) and $P=1$ -- which display the same type of behaviour: a constant real part equal to Log(3)/4$\pi(r_++r_-)$ (in the Schwarzchild case, $r_-=0$ and $r_+=r_s=1$) and an imaginary gap respectfully equal to 1/2 and 1. In between these extremal cases, \textit{i.e} for $0<P<1$, the real part of QNMs is oscillating between 0 and Log(5)/$4\pi(r_+-r_-)$ as one can see in figure \ref{Modesto_mono_plot_s0}.

\begin{figure}[!h]
    \begin{subfigure}[b]{0.42\textwidth}
    \centering  
    \includegraphics[width=1\textwidth]{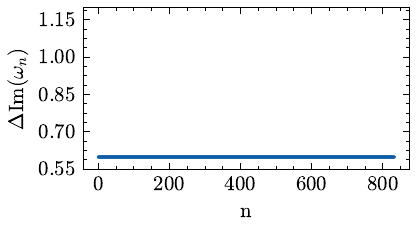}
    \caption{$P=0.1$}
    \end{subfigure}
    \hspace{0.5cm}
    \centering
    \begin{subfigure}[b]{0.42\textwidth}
    \centering  
    \includegraphics[width=1\textwidth]{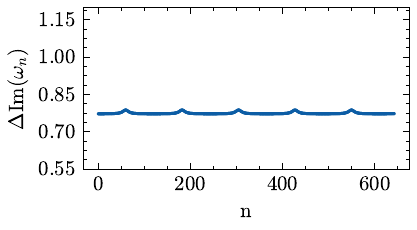}
    \caption{$P=0.3$}
    \end{subfigure}
    \hspace{0.5cm}
    \\
    \centering
    \begin{subfigure}[b]{0.42\textwidth}
    \centering  
    \includegraphics[width=1\textwidth]{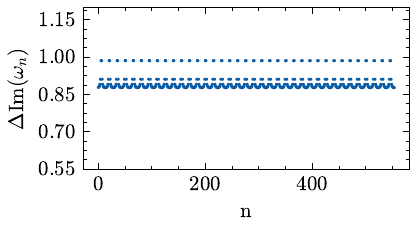}
    \caption{$P=0.5$}
    \end{subfigure}
    \
    \\
    \begin{subfigure}[b]{0.42\textwidth}
    \centering  
    \includegraphics[width=1\textwidth]{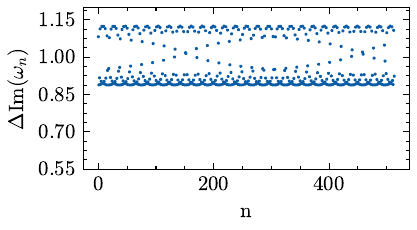}
    \caption{$P=0.7$}
    \end{subfigure}
    \hspace{0.5cm}
    \centering
    \begin{subfigure}[b]{0.42\textwidth}
    \centering  
    \includegraphics[width=1\textwidth]{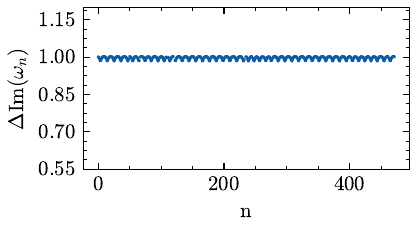}
    \caption{$P=0.9$}
    \end{subfigure}
    \hspace{0.5cm}
    \caption{Plots on the complex plane of the evolution of the imaginary gap $\Delta$Im$(\omega_n)$ in terms of the mode number $n$ for scalar modes computed via the monodromy technique. Five different values of $P$ are represented, from 0.1 to 0.9.}
    \label{Modesto_mono_IM_plot_s0}
\end{figure}

It turns out that the gap in imaginary part also displays some oscillations patterns. Although it is not well visible for the naked eye, it explains the chaotic spectra of $P\sim0.6$ where both oscillations in the real part and in the imaginary gap mix together. These oscillations patterns for the imaginary gap are represented in figure \eqref{Modesto_mono_IM_plot_s0}, and one can clearly see how $\Delta$Im$(\omega_n)$ goes from 1/2 ($P=0$) to 1($P=1)$.

\ \\

\begin{figure}[!h]
    \centering
    \includegraphics[width=0.6\linewidth]{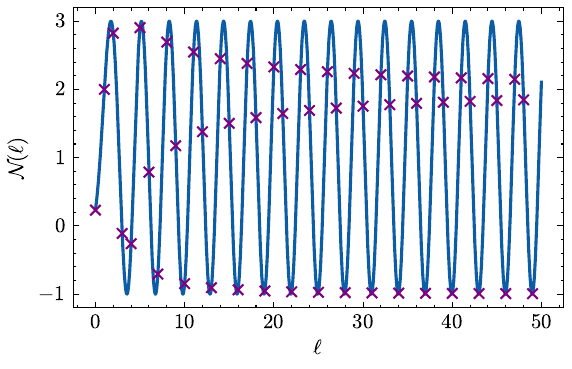}
    \caption{Plot of the coefficient $\mathcal{N}(\ell)=1+2 \cos \left[\frac{\sqrt{4 \ell(\ell+1)+17}}{3}\pi\right]$ for $0<\ell<50$. }
    \label{Modesto_P_plot_coeff_mono_s2}
\end{figure}

Now, let us cover the case of spin $s=2$ test-field perturbations. This time the final equation describing the QNMs computed from the monodromy technique is more involved and depends on the angular momentum parameter $\ell$ via the coefficient $\mathcal{N}(\ell)=1+2 \cos \left[\frac{\sqrt{4 \ell(\ell+1)+17}}{3}\pi\right]$. Indeed, the equation \eqref{mono_eq_s2} can be written as:
\be 
e^{\frac{4 \pi  r_+^2 \omega }{r_+-r_-}}+ e^{\frac{4 \pi r_-^2 \omega }{r_+-r_-}}[1+\mathcal{N}(\ell)]+\mathcal{N}(\ell) = 0,
\label{eq_mono_s2_coeff}
\ee
The QNM spectra will then be highly influenced by the value of $\mathcal{N}(\ell)$, and in order to better understand this behaviour, we can first take a look at the evolution of this coefficient. To do so, we plotted in figure \ref{Modesto_P_plot_coeff_mono_s2} the coefficient $\mathcal{N}(\ell)$ for $0<\ell<50$.

One can see that the coefficient $\mathcal{N}(\ell)$ oscillates between -1 and 3. The integer values of $\ell$ are marked by a cross and the corresponding values of $\mathcal{N}(\ell)$ are distribute in an erratic way between -1 and 3. We recall that the values $\ell=0$ and $\ell=1$ are not physically valid for spin 2 perturbations as the latter do not propagate in these cases. We are then only interested in $\ell\geq2$. Although the values of $\mathcal{N}(\ell)$ seem erratic for small $\ell$, one can notice that they actually converge for large $\ell$ towards two different values: 2 and -1. In this limit, the equation \eqref{eq_mono_s2_coeff} reduces to two distinct equations: 
\be
\left\{
    \begin{array}{l}
        e^{\frac{4 \pi  r_+^2 \omega }{r_+-r_-}}+3 e^{\frac{4 \pi r_-^2 \omega }{r_+-r_-}}+2 = 0, \quad \quad \text{for \ } \mathcal{N}(\ell)=2,\\
       e^{\frac{4 \pi r_+^2 \omega}{r_+-r_-} }-1= 0 \quad \quad \text{for \ } \mathcal{N}(\ell)= -1.
    \end{array}
    \right.
    \label{mono_eq_s2}
\ee
One can notice that these equations are the same than the ones for the spin 0 case. In the limit $\ell\gg1$ the spin 2 QNM spectra predicted by the monodromy calculation will then be the same as the ones we described previously. 

We can verify the behaviour at large $\ell$ of the coefficient $\mathcal{N}(\ell)$:
\be
\mathcal{N}(\ell) \overset{\ell\gg1}{\sim} 1+2\cos\lc \frac{2\ell+1}{3}\pi \rc. 
\ee 
One can then notice that if $\ell$ can be written as $\ell=1+3p, \, p\in\mathbb{N}$ -- \textit{i.e.} if its remainder when divided by 3 is 1 --, then $\mathcal{N}(\ell) = 1+2\cos \lc (1+2p)\pi \rc = -1$. If it if not case, \textit{i.e.} if the remainder of $\ell$ when divided by 3 is 0 or 2, then we obtain that $\mathcal{N}(\ell)=2$. We therefore find ourselves in exactly the same situation as scalar perturbation \eqref{mono_eq_s0} for $\ell\gg1$.
\\ \\

\begin{figure}[!h]
    \centering
    \begin{subfigure}[b]{0.23\textwidth}
    \centering  
    \includegraphics[width=1\textwidth]{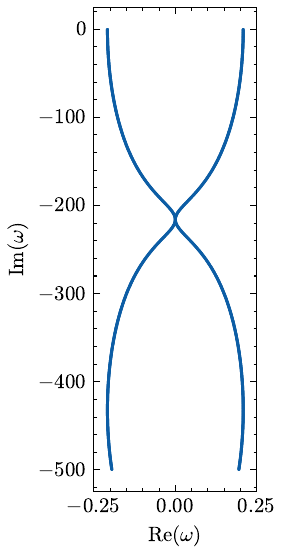}
    \caption{$P=0.2$}
    \end{subfigure}
    \hfill
    \centering
    \begin{subfigure}[b]{0.23\textwidth}
    \centering  
    \includegraphics[width=1\textwidth]{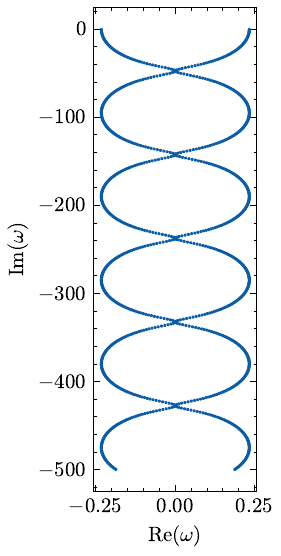}
    \caption{$P=0.3$}
    \end{subfigure}
    \hfill
    \centering
    \begin{subfigure}[b]{0.23\textwidth}
    \centering  
    \includegraphics[width=1\textwidth]{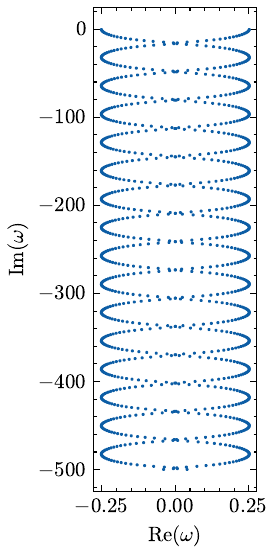}
    \caption{$P=0.4$}
    \end{subfigure}
    \hfill
    \centering
    \begin{subfigure}[b]{0.23\textwidth}
    \centering  
    \includegraphics[width=1\textwidth]{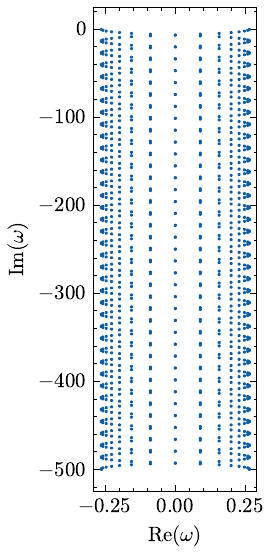}
    \caption{$P=0.5$}
    \end{subfigure}
    \hfill
    \centering
    \begin{subfigure}[b]{0.23\textwidth}
    \centering  
    \includegraphics[width=1\textwidth]{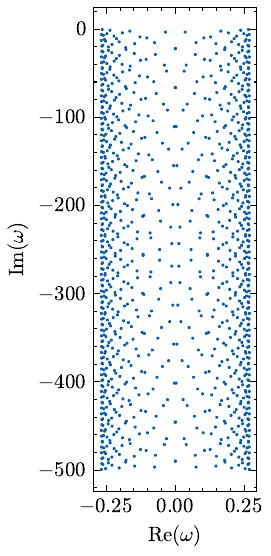}
    \caption{$P=0.6$}
    \end{subfigure}
    \hfill
    \centering
    \begin{subfigure}[b]{0.23\textwidth}
    \centering  
    \includegraphics[width=1\textwidth]{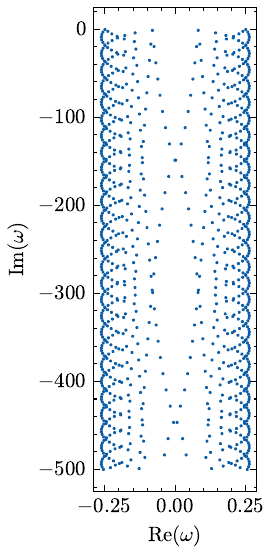}
    \caption{$P=0.7$}
    \end{subfigure}
    \hfill
    \centering
    \begin{subfigure}[b]{0.23\textwidth}
    \centering  
    \includegraphics[width=1\textwidth]{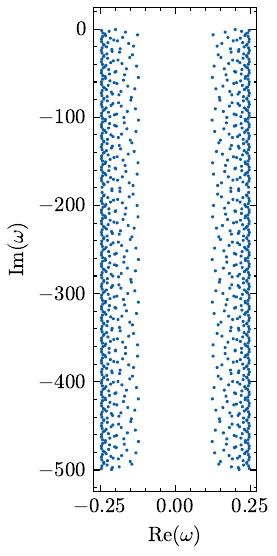}
    \caption{$P=0.8$}
    \end{subfigure}
    \hfill
    \centering
    \begin{subfigure}[b]{0.23\textwidth}
    \centering  
    \includegraphics[width=1\textwidth]{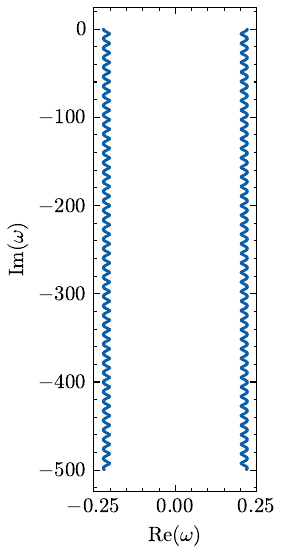}
    \caption{$P=0.9$}
    \end{subfigure}
    \hfill    
    \caption{Plots on the complex plane of the QNMs values predicted by the monodromy calculation, for spin $s=2$ test field perturbations  and $\ell=2$.}
    \label{Modesto_mono_plot_s2}
\end{figure}
\begin{figure}[!htbp]
    \hspace{0.5cm}
    \begin{subfigure}[b]{0.26\textwidth}
    \centering  
    \includegraphics[width=1\textwidth]{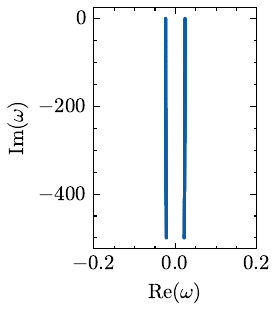}
    \caption{$\ell=3, \ P=0.1$}
    \end{subfigure}
    \hspace{0.5cm}
    \begin{subfigure}[b]{0.25\textwidth}
    \centering  
    \includegraphics[width=1\textwidth]{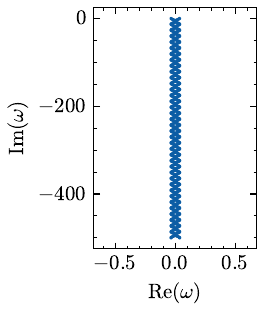}
    \caption{$\ell=3, \ P=0.5$}
    \end{subfigure}
    \hspace{0.5cm}   
    \begin{subfigure}[b]{0.25\textwidth}
    \centering  
    \includegraphics[width=1\textwidth]{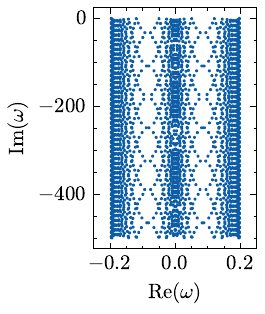}
    \caption{$\ell=3, \ P=0.9$}
    \end{subfigure}
    \hspace{1.5cm}
    \ \\
    \hspace{0.5cm}
    \begin{subfigure}[b]{0.26\textwidth}
    \centering  
    \includegraphics[width=1\textwidth]{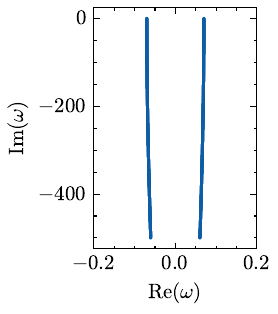}
    \caption{$\ell=4, \ P=0.1$}
    \end{subfigure}
    \hspace{0.5cm}
    \begin{subfigure}[b]{0.25\textwidth}
    \centering  
    \includegraphics[width=1\textwidth]{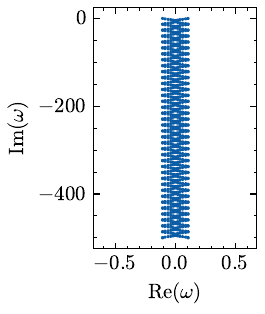}
    \caption{$\ell=4, \ P=0.5$}
    \end{subfigure}
    \hspace{0.5cm}    
    \begin{subfigure}[b]{0.25\textwidth}
    \centering  
    \includegraphics[width=1\textwidth]{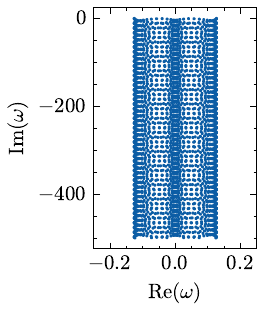}
    \caption{$\ell=4, \ P=0.9$}
    \end{subfigure}
    \hspace{1.5cm}
    \ \\
    \hspace{0.5cm}
    \begin{subfigure}[b]{0.26\textwidth}
    \centering  
    \includegraphics[width=1\textwidth]{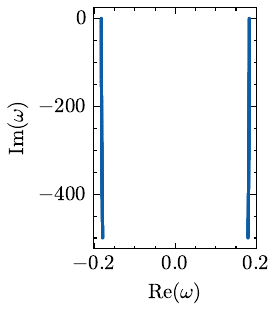}
    \caption{$\ell=5, \ P=0.1$}
    \end{subfigure}
    \hspace{0.5cm}
    \begin{subfigure}[b]{0.25\textwidth}
    \centering  
    \includegraphics[width=1\textwidth]{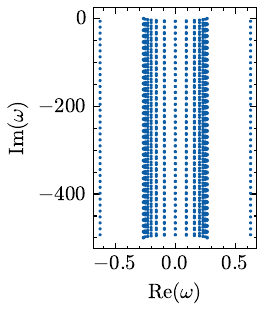}
    \caption{$\ell=5, \ P=0.5$}
    \end{subfigure}
    \hspace{0.5cm}   
    \begin{subfigure}[b]{0.25\textwidth}
    \centering  
    \includegraphics[width=1\textwidth]{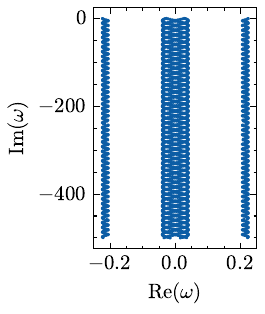}
    \caption{$\ell=5, \ P=0.9$}
    \end{subfigure}
    \hspace{1.5cm}
 
    \caption{Plots on the complex plane of the QNMs values predicted by the monodromy calculation for spin $s=2$ test field perturbations. Each line gives a different value of the angular momentum parameter $\ell \, (3\leq\ell\leq5)$ and each column a different value of $P \, (P=0.1, \, 0.5, \, 0.9)$.}
    \label{Modesto_mono_plot_s2_several_l}
\end{figure}
\begin{figure}[!htbp]
    \hspace{0.5Cm}
    \begin{subfigure}[b]{0.25\textwidth}
    \centering  
    \includegraphics[width=1\textwidth]{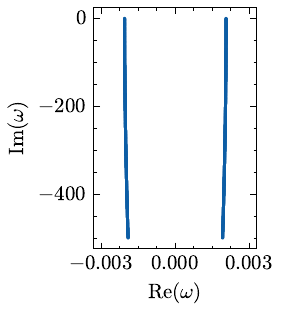}
    \caption{$P=0.1$}
    \end{subfigure}
    \hspace{0.5cm}
    \centering
    \begin{subfigure}[b]{0.25\textwidth}
    \centering  
    \includegraphics[width=1\textwidth]{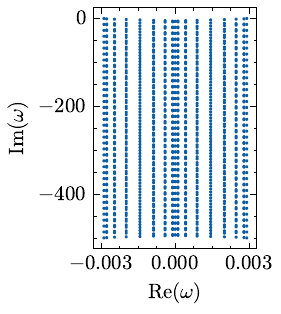}
    \caption{$P=0.5$}
    \end{subfigure}
    \hspace{0.5cm}
    \begin{subfigure}[b]{0.25\textwidth}
    \centering  
    \includegraphics[width=1\textwidth]{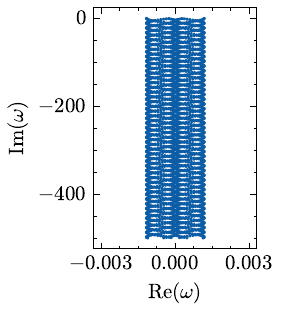}
    \caption{$P=0.9$}
    \end{subfigure}
    \hspace{1.5cm}
\caption{Plots on the complex plane of the QNMs values predicted by the monodromy calculation for spin $s=2$ test field perturbations, $\ell=40$ and $P=0.1, 0.5, 0.9$.}
\label{Modesto_mono_plot_s2_l40}
\end{figure}

For small $\ell$, the equations will differ however, and the same applies to the related QNM spectra. To gain an understanding of these changes, we displayed the spectra predicted by the monodromy calculation for different values of the angular momentum parameter $\ell$ and several values of $P$. 
\\
Let us start by $\ell=2$, the lowest possible value of $\ell$ for spin 2 perturbations. The QNM spectra for $\ell=2$ and $0.2\leq P \leq 0.9$ are shown in figure \ref{Modesto_mono_plot_s2}. One can see that the spectra are actually pretty similar from the ones we obtain for scalar perturbations (see figure \ref{Modesto_mono_plot_s0}). The only notable difference lies in the amplitudes of the oscillations: one can notice that the maxima have slightly different values.  
\\
The spectra show more variations as we increase $\ell$ nonetheless. We displayed the QNM spectra for $\ell=3, \, 4, \, 5$ and for three values of $P$ in figure \ref{Modesto_mono_plot_s2_several_l}. One can see that both the amplitude and the patterns of the oscillations change depending on the value of $\ell$.
\\
The QNM spectra will continue to differ when increasing the angular momentum parameter $\ell$ until we will reach the limit $\ell\gg 1$ where either $\mathcal{N}(\ell)=2$ or $\mathcal{N}(\ell)=-1$. In the latter case, the real part of the QNMs predicted by the monodromy calculation vanish and the QNMs are then purely damped. In order to gain an insight of the transition to this limit, we can take a look at the case $\ell=40$ for which $\mathcal{N}(\ell)\gtrsim -1$ as one can see on figure \ref{Modesto_P_plot_coeff_mono_s2}. The plots for $P= 0.1, \, 0.5, \, 0.9$ are shown in figure \eqref{Modesto_mono_plot_s2_l40}. One can see that the maximal amplitude of the real part is about $10^{-3}$ and that we are indeed in the midst of a transition towards Re$(\om)=0$.

We have not discussed the behaviour of the Imaginary gap for the case of spin 2 test-field perturbations. In short, the amplitudes and the patterns of the oscillations will as well be diverse for small values of $\ell$, and will then either settle to a constant imaginary gap (see equation \eqref{special_Im}) in the case $\mathcal{N}(\ell)=-1$, or will behave as what we observed in figure \ref{Modesto_mono_IM_plot_s0} in the case $\mathcal{N}(\ell)=2$.

\subsection{Continued fraction results}

%
\begin{figure}[!htbp]
    \begin{subfigure}[b]{0.26\textwidth} 
    \centering
    \includegraphics[width=1\textwidth]{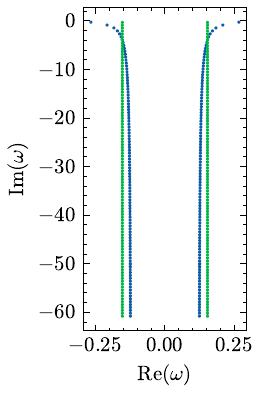}
    \caption{$P=0.1$}
    \end{subfigure}
    \hspace{0.5cm}
    \begin{subfigure}[b]{0.25\textwidth}
    \centering  
    \includegraphics[width=1\textwidth]{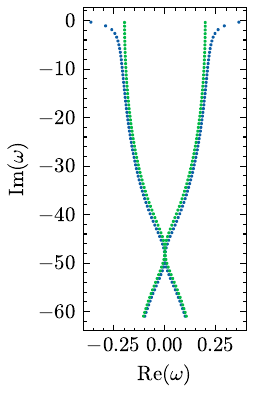}
    \caption{$P=0.3$}
    \end{subfigure}
    \hspace{0.5cm}  
    \begin{subfigure}[b]{0.26\textwidth}
    \centering
    \includegraphics[width=1\textwidth]{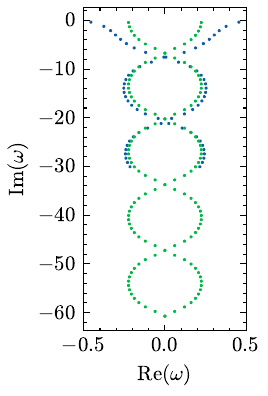}
    \caption{$P=0.5$}
    \end{subfigure}
    \hspace{2cm}
    \centering
    \begin{subfigure}[b]{0.27\textwidth}
    \centering  
    \includegraphics[width=1\textwidth]{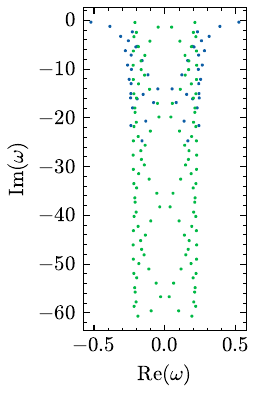}
    \caption{$P=0.7$}
    \end{subfigure}
    \hspace{0.5cm}
    \begin{subfigure}[b]{0.27\textwidth}
    \centering  
    \includegraphics[width=1\textwidth]{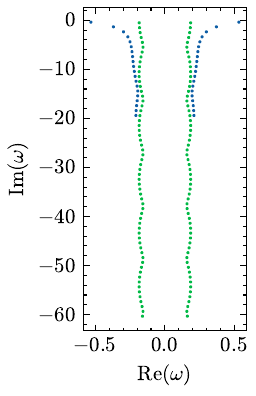}
    \caption{$P=0.9$}
    \end{subfigure}
    \hspace{1cm}
\caption{Plots on the complex plane of the QNMs values predicted by the Continued fraction method (in \textcolor[HTML]{0C5DA5}{blue}) and by the monodromy calculation (in \textcolor[HTML]{00B945}{green}) for scalar field perturbations, $\ell=s=0$ and $P=0.1, \, 0.3, \, 0.5, \, 0.7, \, 0.9$.}
\label{Modesto_leaver_mono_plot_s0}
\end{figure}

In this section we will now present our results from the Continued fraction method and directly compare them the ones we obtained from the monodromy method. \\
For the scalar perturbations, we show in figure \ref{Modesto_leaver_mono_plot_s0} five plots of QNM spectra for $\ell=0$ and five different values of $P$ between 0.1 and 0.9. The QNMs computed with Leaver's method are shown in blue while the ones computed via the monodromy technique are represented in green. One can first notice that the larger $P$ is and the more oscillation patterns there are, the less number of QNMs we were able to compute via the Continued fraction method. While we could easily compute QNMs with $|$Im$(\om)|\gtrsim 100$ for $P\sim0.1$, we could only reach $|$Im$(\om)|\sim 20$ for $P\gtrsim0.7$. However, one can see that the monodromy and the Leaver spectra show very similar behaviour for each value of $P$. In particular, for $P=0.3$, the values of both types of QNMs seem to be getting closer and closer as their imaginary part increases. For $P=0.1$, we find ourselves in the classic case of vertical asymptote and we can then apply a fit on both types of asymptotic spectra in order to compare the predictions of asymptotic real part and imaginary gap. 
The results of the fits on both spectra yield:

\be
\left\{
    \begin{array}{l}
        \text{Re}(\om_n^{\text{(mono)}}) = (0.1516\pm 0.0003) + \frac{(0.012\pm 0.005)}{\sqrt{n}},\\
       \Delta\text{Im}(\om_n^{\text{(mono)}}) = (0.5989858\pm 10^{-7}),
    \end{array}
    \right.
    \label{mono_pred_P01}
\ee
for the monodromy one and 
\be
\left\{
    \begin{array}{l}
        \text{Re}(\om_n^{\text{(leaver)}}) = (0.122\pm 0.001) + \frac{(0.03\pm 0.01)}{\sqrt{n}},\\
       \Delta\text{Im}(\om_n^{\text{(leaver)}}) = (0.597\pm 0.009) + \frac{(0.03\pm 0.05)}{\sqrt{n}},
    \end{array}
    \right.
    \label{}
\ee
for the one computed from the Leaver's method. 
\\
The corresponding relative errors between the value predicted by the monodromy technique and the value computed from the Continued fraction method are then 19.5$\%$ for the Real part and 0.33$\%$ for the Imaginary gap. The latter one is very small, as expected for the Imaginary part, and also confirms the conjecture done in the previous study -- $\Delta$Im$(\om) \sim 0.5 + P$, \cite{Livine:2024bvo} -- is accurate for small values of $P$. 

\begin{figure}[!htbp]
    \centering
    \includegraphics[width=0.5\linewidth]{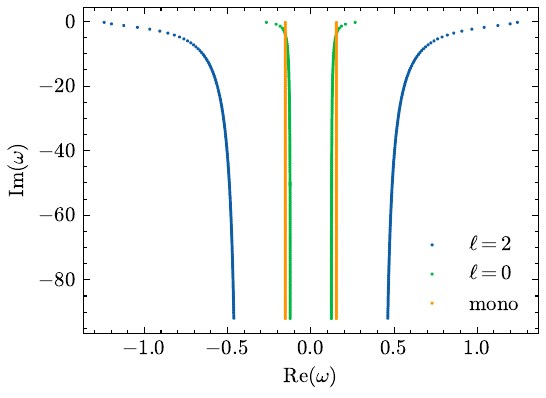}
    \caption{Plot on the complex plane of the QNMs values predicted by the Continued fraction method for scalar field perturbations, $\ell=0$ and $\ell=2$, and $P=0.1$. We plotted as well the related monodromy prediction. }
    \label{plot_l02_P01}
\end{figure}

Following this, we can take a look at the results given by the Continued fraction method for $\ell=2$, which are supposed to have the same asymptotic behaviour as $\ell=0$ according to the monodromy prediction (see \eqref{mono_eq_s0}). We specifically considered the value $P=0.1$, as it is easier to compare the asymptotic of a Schwarzschild-type spectrum. One can find on figure \ref{plot_l02_P01} plots of QNMs with $\ell=0$ and $\ell=2$ for $P=0.1$, and as well the corresponding monodromy prediction for comparison. While the spectrum for $\ell=0$ seems relatively close to the monodromy one, the spectrum for $\ell=2$ shows larger real part. We performed a fit in order to have a more precise idea. We found that, for $\ell=2$ and $P=0.1$:
\be
\left\{
    \begin{array}{l}
        \text{Re}(\om_n^{\text{(leaver)}}) = (0.42\pm 0.06) + \frac{(1.0\pm 0.5)}{\sqrt{n}} + \frac{12\pm 5}{n},\\
       \Delta\text{Im}(\om_n^{\text{(leaver)}}) = (0.598\pm 0.001) + \frac{(0.01\pm 0.02)}{\sqrt{n}}.
    \end{array}
    \right.
    \label{}
\ee
\begin{figure}[!htbp]
    \centering
    \includegraphics[width=0.5\linewidth]{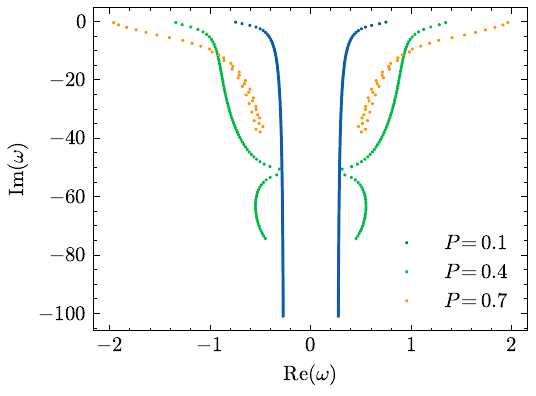}
    \caption{Plots on the complex plane of the QNMs values predicted by the Continued fraction method for scalar field perturbations, $\ell=1$ and $P=0.1, \, 0.4, \, 0.7$.}
    \label{plot_QNMs_P_s0_l1}
\end{figure}
\begin{figure}[!htbp]
    \centering
    \includegraphics[width=0.55\linewidth]{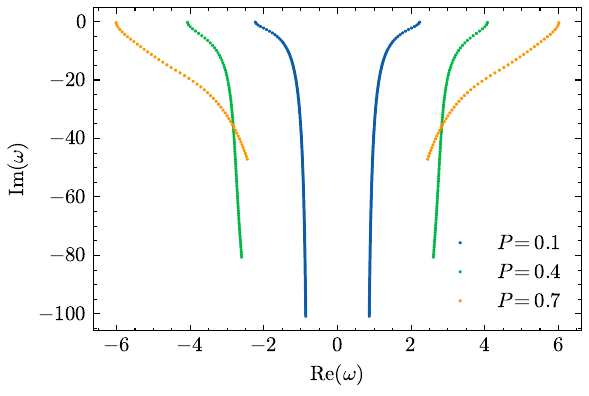}
    \caption{Plots on the complex plane of the QNMs values predicted by the Continued fraction method for scalar field perturbations, $\ell=4$ and $P=0.1, \, 0.4, \, 0.7$.}
    \label{plot_QNMs_P_s0_l4}
\end{figure}
\begin{figure}[!htbp]
    \begin{subfigure}[b]{0.26\textwidth}
    \centering  
    \includegraphics[width=1\textwidth]{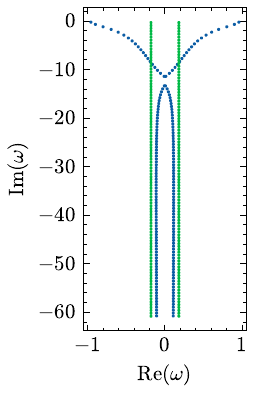}
    \caption{$P=0.1$}
    \end{subfigure}
    \hspace{0.5cm}
    \centering
    \begin{subfigure}[b]{0.26\textwidth}
    \centering  
    \includegraphics[width=1\textwidth]{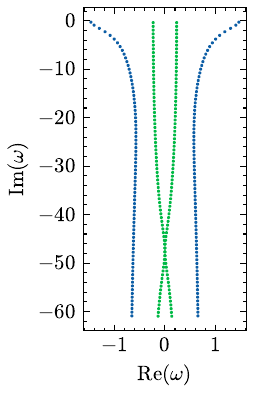}
    \caption{$P=0.3$}
    \end{subfigure}
    \hspace{0.5cm}  
    \begin{subfigure}[b]{0.26\textwidth}
    \centering  
    \includegraphics[width=1\textwidth]{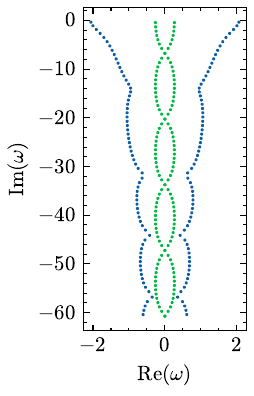}
    \caption{$P=0.5$}
    \end{subfigure}
    \hspace{1cm}
    \begin{subfigure}[b]{0.27\textwidth}
    \centering  
    \includegraphics[width=1\textwidth]{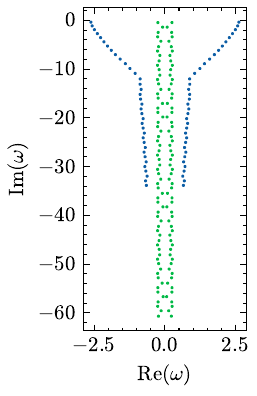}
    \caption{$P=0.7$}
    \end{subfigure}
    \hspace{0.5cm}
    \begin{subfigure}[b]{0.27\textwidth}
    \centering  
    \includegraphics[width=1\textwidth]{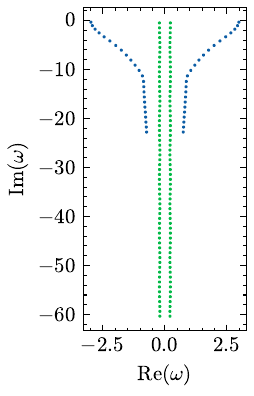}
    \caption{$P=0.9$}
    \end{subfigure}
\caption{Plots on the complex plane of the QNMs values predicted by the Continued fraction method (in \textcolor[HTML]{0C5DA5}{blue}) and by the monodromy calculation (in \textcolor[HTML]{00B945}{green}) for spin 2 test-field perturbations, $\ell=s=2$ and $P=0.1, \, 0.3, \, 0.5, \, 0.7, \, 0.9$.}
\label{Modesto_leaver_mono_plot_s2}
\end{figure}

One can notice that the result for the Imaginary gap is in very good agreement with the monodromy prediction \eqref{mono_pred_P01}. However, it is not the case for the Real part, which is almost three times larger than the monodromy one. Yet, it is not that surprising if we compare to the Schwarzschild case. Indeed, it is known that the larger the angular parameter is, the greater its influence on the asymptotic value. Therefore, the larger $\ell$ is, the longer the QNMs will take to achieve the asymptote, and it is needed to compute an even large number of QNMs to see it \cite{Nollert:1993zz}.
\\

Now, let us take a look at QNM spectra with values of $\ell$ for which the monodromy predicts purely damped QNMs (\textit{i.e.} with vanishing real part, see equation \eqref{mono_eq_s0}). It is the case for $\ell=1$ and $\ell=4$. One can see the corresponding spectra for different values of $P$ in figures \ref{plot_QNMs_P_s0_l1} and \ref{plot_QNMs_P_s0_l4} respectfully.

One can see that, in the range we were able to compute QNMs with the continued fraction method, the real part of the modes does not seem to tend to zero. It is difficult to say if these QNMs will actually achieve this asymptotic behaviour at larger imaginary part, or if they will display a different asymptote than the one predicted by the monodromy calculation.
Indeed, as we already mentioned, the more the angular momentum parameter $\ell$ is large, the later the asymptote is reached and it becomes difficult to witness it from the Continued fraction method only.

To finish, we can take a look at the test-field spin 2 QNMs computed via the Continued fraction method, and compare them to the monodromy prediction. We display the results for $\ell=s=2$ and five different values of $P$ in figure \ref{Modesto_leaver_mono_plot_s2}, along with the monodromy spectra from figure \ref{Modesto_mono_plot_s2}. The first thing to be noticed is that the well-known crossing of the imaginary axis of gravitational Schwarzschild QNMs seem to disappear from $P\gtrsim0.3$. There is then an important structure shift in the QNM spectra when $P$ is increased. The second thing one can notice is that the QNMs computed via the Continued fraction method are less matching with the monodromy prediction than the ones for $s=\ell=0$ (see figure \ref{Modesto_leaver_mono_plot_s0}). Yet, the real part of the Continued fraction QNMs is decreasing as the imaginary part increases, and gets closer to the monodromy one. Moreover, the period of the oscillations for $P=0.5$ and $P=0.7$ seem to converge to the one of the monodromy QNMs, which gives hope that the asymptote of the Continued fraction QNMs indeed corresponds to the monodromy one. And once again, it is known the gravitational QNMs of the Schwarzschild BH reach the asymptote later than the scalar ones. The same phenomena could apply here.

\section*{\uppercase{Discussion and conclusion}}

Quasi-Normal Modes (QNMs) of Black Holes (BHs) not only allow to test the stability of BHs, they also enable to characterise one BH and study the effect of its parameter. Since the first detection of GWs in 2015, they are as well seen as a great tool to test GR. 

In this study, we chose to move away from this perspective by studying perturbations of an effective BH model which cannot be related to the measurement of Gravitational Waves (GWs), and as well by spanning the hole range of possible values of the BH parameter, going well over the astrophysical predictions. These choices are motivated by the desire to perform a mathematical study of the QNMs, by analysing the deviations from GR and the symmetries which arise or vanish. 
The BH we study here is one of the first effective model that has been derived following the principles of Loop Quantum Gravity (LQG).
It was built by Modesto in 2008 \cite{Modesto:2008im} and implements the Loop quantum deviations via two parameters: the area gap $a_0$ resolving the singularity issue and the polymeric function $P$ modifying the horizon structure. As we already stated before, proper gravitational perturbations of effective models of BH as this one cannot be studied as we are lacking equations describing the dynamic of space-time in LQG. The link with GWs detection cannot then be done. What is possible to do, however, is to study spin 0 and 2 test-field perturbations which will account for the deviations in the BH structure itself and allow to characterise it. 

With this perspective in mind, we chose to not restrict ourselves to low damped QNMs as in our previous study, but rather explore the hole spectra of QNMs. To do so, we combined two different methods of calculation of QNMs. The first one is the one we already used in our previous study: the Continued fraction method. Its high accuracy, especially for the low and mid damped QNMs has then been verified. However, this method does not converge properly for modes with a large imaginary part, especially if there are oscillating patterns. This is why we have been willing to use, as complement, another method which is especially made for highly damped modes: the monodromy technique. 
The case of Modesto BH is quite close from the Reissner-Nordström case, and the monodromy technique allows to highlight the existence of asymptotic oscillating patterns. 

We chose here to restrict ourselves to the study of the impact of the parameter $P$ as the monodromy technique for $a_0$ is more involved (see appendix \ref{appendix_mono_Modesto_a0_case}). Moreover, $a_0$ is supposed to be, by construction, of Planck order while deviations from the Schwarzschild BH can only be \enquote{seen} from $a_0\sim 10^{-1}$ (see \cite{Livine:2024bvo}). 
\\
In our last study, we only focused on small values of $P$. Here, we chose to study the hole range of physical values $0<P<1$ and then discovered the oscillating patterns, both in the Real part and in the Imaginary gap. We highlighted and described the evolution of these phenomena with the polymeric function $P$, the angular momentum parameter $\ell$ and for both spin 0 and spin 2 test-field perturbations. 
\\
Although the Continued fraction method showed weakness in the computation of oscillating and highly damped QNMs, it still allowed to compute QNMs on a large part of the complex spectrum. It moreover showed great consistency with the results of the monodromy technique, which predicts the asymptotic behaviour. The combination of these two techniques allowed us to analyse efficiently the QNMs on the hole complex plane. 

The study of highly damped QNMs brings us back to the question of their meaning. If the constant Imaginary gap of Schwarzschild BH is truly related to the discrete area spectrum of LQG \cite{Dreyer:2002vy}, then how to explain the deviations observed in the case of this effective model derived from LQG? The question arises all the more in view of the oscillations patterns appearing. Which physical meaning do these asymptotic oscillations have compared to the straight asymptote of the Schwarzschild BH?

\appendix

\section{Monodromy technique for the case $\ell=1$ ($a_0=0$)}
\label{appendix_mono_Modesto_P_case_l1}

In this section we consider the case $\ell=1$ for the computation of asymptotic scalar QNMs via the monodromy technique. The core idea of this technique is to solve the master equation describing the evolution of the perturbations around $r=0$, in order to compute the monodromy of the field function around a well-chosen contour. To do so, we use the first order expression of the effective potential around $r=0$, which reads in the case of scalar perturbations on the Modesto BH (where we set again $a_0=0$):
\be 
V_0(r)\overset{r=0}{\sim} (\lambda-2)\frac{P^4}{(1+P)^8}\frac{1}{r^6}+\mathcal{O}(1).
\ee
However, as $\lambda=\ell(\ell+1)$, one can see that this expression vanishes for $\ell=1$. The final equation \eqref{mono_Modesto_P_result} cannot then be valid for the case $\ell=1$. In order to handle this special case correctly, one has to use next-to-leading order for the effective potential:
\be 
V_0^{(\text{NTLO})}(r)\overset{r=0}{\sim} (3-\lambda)\frac{P^2}{(1+P)^6}\frac{1}{r^5}+\mathcal{O}(2),
\label{NTLO_pot_Modesto_P}
\ee
which is actually the leading order in the case $\ell=1$.
\\
Using the first order approximation of tortoise coordinate $x(r)$ around $r=0$,
\be 
x\overset{r=0}{\sim} \frac{(1+P)^4}{3P^2}r^3,
\ee 
we can write the NTLO effective potential \eqref{NTLO_pot_Modesto_P} in terms of the tortoise coordinate $x$:
\be 
V_0^{(\text{NTLO})}(r)\overset{r=0}{\sim} \frac{3-\lambda}{3^{5/3}}\frac{(1+P)^{2/3}}{P^{4/3}}\frac{1}{x^{5/3}} \equiv \frac{\mathcal{V}}{x^{5/3}}.
\ee
The master equation describing scalar perturbations around the Modesto BH can then be simplified around $r=0$ for the case $\ell=1$ as
\be
\frac{\dd^2 \Psi}{\dd x^2}+[\omega^2-\frac{\mathcal{V}}{x^{5/3}}]\Psi = 0.
\ee
We can introduce another function $\Phi$ such that $\Psi(x)=\sqrt{2\pi \omega x} \ \Phi(\omega x)$, in such a way that the function $\Phi$ satisfies the equation 
\be 
\omega^2x^2\Phi''(\omega x) + \omega x \Phi'(\omega x) + \lc \omega^2x^2 - \mathcal{V}x^{1/3} - \frac{1}{4} \rc \Phi(\omega x) = 0.
\ee 
However, this equation does not correspond to a Bessel equation because of the term in $x^{1/3}$, and we were not able to find any other function satisfying this equation.

\section{Monodromy technique for the case ($a_0\neq0,P=0$)}
\label{appendix_mono_Modesto_a0_case}

\begin{figure}[!htbp]
     \centering
     \begin{subfigure}[b]{0.45\textwidth}
         \centering
         \includegraphics[width=0.95\textwidth]{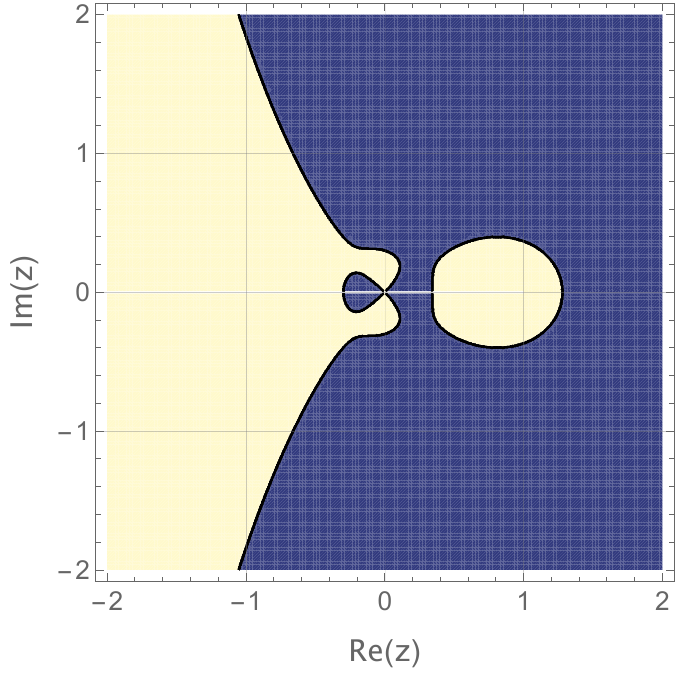}
         \caption{$r_0=0.1$}
     \end{subfigure}
     \hspace{0.2cm}
     \begin{subfigure}[b]{0.45\textwidth}
         \centering
         \includegraphics[width=0.95\textwidth]{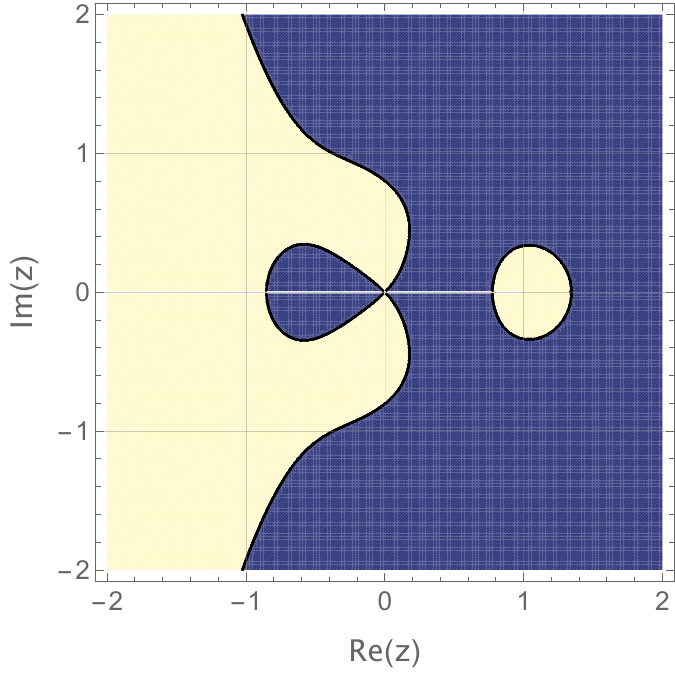}
         \caption{$r_0=0.9$}
     \end{subfigure}
     \caption{Complex $r$-plane with Stokes lines represented in black for the tortoise coordinate $x$ defined \textit{via} equation \eqref{tort_a0}, for $a_0=0.1$ and $a_0=0.9$ (with $r_s=1$). The dark blue regions correspond to Re($x) > 0$.}
     \label{r_planes_a0}
\end{figure}

In this section we will consider the case of the Modesto BH with $P=0$ and $a_0$ being non-zero. The numerical computation of QNMs in this case has already been considered our previous study \cite{Livine:2024bvo}, revealing interesting patterns. Here we want to examine the monodromy technique in this context, which could be useful to study the asymptotic spectrum in $a_0$.\\
First, let us recall the metric functions in the case $P=0$. The line element of the Modesto BH is given in equation \eqref{Modesto_new_metric} and the metric reads:
\be 
f(r) = \frac{(r-r_s) \, r^3}{a_0^2+r^4}, \quad g(r) = \frac{(r-r_s) \, r^3}{a_0^2+r^4}, \quad h(r) = r^2+\frac{a_0^2}{r^2}.
\label{metric_a0}
\ee
This metric then represent a singularity free BH, with a minimal area given by the parameter $a_0$ which is then expected to be of Planck order. 
\\
As the Modesto BH is a static and spherically BH, we can once again directly use the master equation for the field function $\Psi(r)$
\be
\frac{\dd^2 \Psi}{\dd x^2}+[\omega^2-V_s(r)]\Psi = 0,
\ee
which controls the evolution of the test-field perturbations.\\
The effective potentials $V_s(r)$ are given by the general formulas for scalar and test-field spin 2 perturbations applied to the metric functions \eqref{metric_a0} and then read:
\begin{subequations}
    \begin{align}
        V_s(r) = \frac{(r-r_s)r^4}{(a_0^2+r^4)^4} \left[ \lambda r^9 + (1-2s)r_sr^8 + 2a_0^2 (\lambda+5-7s)r^5 + (12s-10)a_0^2r_sr^4 \right. \nonumber \\ \left.+ a_0^4(\lambda-2+2s)r + (1-2s)a_0^4 r_s \right]. 
    \end{align}
\end{subequations}
On the other side, the tortoise coordinate $x$ reads:
\be 
x(r) = r+ \frac{a_0^2 (2 r+r_s)}{2 r_s^2 r^2}-\frac{a_0^2 \log (r)}{r_s^3}+\frac{\left(a_0^2+r_s^4\right) \log (r-r_s)}{r_s^3}.
\label{tort_a0}
\ee 
As we already explain, the monodromy technique focuses on the behaviour of the master equation around $r=0$. First we can take a look at the effective potential's conduct. Its first order contribution around $r=0$ reads
\be 
V_s(r) \overset{r=0}{\sim} (2s-1) \frac{r_s^2 r^4}{a_0^4} + \mathcal{O}(1).
\ee 
Moreover, the first order term of the tortoise coordinate is
\be 
x(r) \overset{r=0}{\sim} \frac{a_0^2}{2r_s r^2} + \mathcal{O}(1),
\ee 
such that the first order contribution of the effective potential $V_s(r)$ can be written in terms of the tortoise coordinate: 
\be 
V_s(r) \overset{r=0}{\sim} \frac{2s-1}{4} \frac{1}{x^2} + \mathcal{O}(1).
\ee 
The master equation then can be written simply around $r=0$ as
\be 
\frac{\dd^2 \Psi}{\dd x}^2 + \lp\omega^2-\frac{2s-1}{4x^2}\rp \Psi  =0.
\ee
As usual, one can rescale the $x$-coordinate into $z=\om x$ and introduce the function $\Psi$ such that $\Psi(x)=\Phi(\omega x)\,{\sqrt{2\pi\omega x}}$. This function now satisfies a straightforward Bessel equation:
\be 
z^2\partial_z^2\Phi+z\partial_z\Phi+\lp z^2-\frac{s}{2} \rp\Phi=0,
\ee 
which can then be treated as usual in the process of the monodromy technique.\\
However, in order to apply properly the method, one need to choose a contour following the Stokes lines for the computation of the monodromy of the field function. The behaviour of the tortoise coordinate is represented in the complex $r$-plane in figure \ref{r_planes_a0} with the Stokes line shown as black lines. 
\\
One can see that the light regions (corresponding to Re$(x)<0$ are separated in two distinct regions with no connexion via the Stokes lines. It is therefore difficult to conceive a contour following the Stokes lines in the way of the other cases we covered and this case would require a more in-depth analysis.

\bibliographystyle{bib-style}
\bibliography{biblio.bib}

\end{document}